\documentclass[useAMS,usenatbib,aas_macros]{mn2e}

\usepackage{epsfig}
\usepackage{lscape,graphicx,colortbl}
\usepackage{amssymb}
\usepackage{natbib}
\usepackage{times}
\usepackage{aas_macros}

\newcommand{\msun}{M_\odot}
\newcommand{\ifm}[1]{\relax\ifmmode#1\else$\mathsurround=0pt#1$\fi}
\newcommand{\kms}{\ifmmode\,{\rm km}\,{\rm s}^{-1}\else km$\,$s$^{-1}$\fi}

\newcommand{\hmsun}{\,\ifm{h^{-1}}{M_{\odot}}}
\def\omm{\Omega_{\rm m}}
\def\oml{\Omega_{\Lambda}}

\newcommand{\be}{\begin{equation}}
\newcommand{\ee}{\end{equation}}
\newcommand{\bea}{\begin{eqnarray}}
\newcommand{\eea}{\end{eqnarray}}
\newcommand{\z}{\emph{z}}

\def\m{{\bf m}}
\def\A{{\bf A}}
\def\B{{\bf B}}
\def\ms{m_{\rm star}}
\def\mc{m_{\rm cold}}
\def\mh{m_{\rm hot}}
\def\dotms{\dot{m}_{\rm star}}
\def\dotmc{\dot{m}_{\rm cold}}
\def\dotmh{\dot{m}_{\rm hot}}
\def\fs{f_{\rm s}}
\def\fe{f_{\rm e}}
\def\fr{f_{\rm re}}
\def\fsdb07{f_{\rm s,D}}
\def\fc{f_{\rm c}}
\def\ffd{f_{\rm d}}
\def\fca{f_{\rm ca}}
\def\fha{f_{\rm ha}}
\newcommand{\HI}{H\,\textsc{i}}
\newcommand{\fof}{{\scshape fof~}}
\newcommand{\tune}{\cellcolor[gray]{0.7}}

\begin{document}

\title[The Degeneracy of Galaxy Formation Models]
      {The Degeneracy of Galaxy Formation Models}

\author[E. Neistein \& S. M. Weinmann]
{Eyal Neistein\thanks{E-mails:$\;$eyal@mpa-garching.mpg.de,
simone@mpa-garching.mpg.de}
 \& Simone M. Weinmann$^{\star}$ \\
 Max-Planck-Institut f\"{u}r Astrophysik, Karl-Schwarzschild-Str. 1,
 85748 Garching, Germany}


\date{}
\pagerange{\pageref{firstpage}--\pageref{lastpage}} \pubyear{2009}
\maketitle

\label{firstpage}


\begin{abstract}
We develop a new formalism for modeling the formation and evolution
of galaxies  within  a  hierarchical  universe.  Similarly  to
standard semi-analytical   models   we   trace  galaxies   inside
dark-matter merger-trees  which are  extracted from  a large
$N$-body simulation. The formalism includes treatment of feedback,
star-formation, cooling, smooth  accretion, gas  stripping  in
satellite galaxies,  and merger-induced  star bursts.
However, unlike  in other models, each process  is assumed to have  an
efficiency which  depends only on the  host halo  mass  and
redshift.  This  allows us  to describe  the various components of
the model in a simple and transparent way.
By allowing the efficiencies to have any value for a given halo mass
and redshift, we can easily encompass a large range of scenarios. To demonstrate this
point, we examine several different galaxy formation models, which are all consistent
with  the  observational  data.  Each  model  is  characterized  by
a different  unique feature:  cold accretion  in low  mass haloes,
zero feedback, stars  formed only in merger-induced  bursts, and
shutdown of star-formation after mergers.
Using these  models we are able  to examine the  degeneracy inherent in
galaxy  formation models, and  look for  observational data  that
will help to break this degeneracy. We  show that the full distribution
of star-formation rates in  a given  stellar mass  bin is promising in constraining  the models.
We compare our  approach in detail to the  semi-analytical model of
De Lucia \&  Blaizot. It is shown  that our formalism is  able to
produce a very similar population of  galaxies once the same median
efficiencies per halo mass and redshift  are being used.
Consequently, our approach may  be useful for comparing various
published  semi-analytical models within the same framework.
We provide a public version of the model galaxies on our web-page,
along with a tool for running models with user-defined parameters. Our model
is able to provide results for a 62.5 $h^{-1}$ Mpc box within just a few seconds.
\end{abstract}




\section{Introduction}
\label{sec:intro}

The physics of galaxy formation combines a wide range of
astrophysical phenomena.  On one  side it  relies heavily  on
cosmology  and  on the hierarchical formation  of dark matter
structures in our Universe. On  the other hand galaxy formation  is greatly
affected by the  detailed physics of gas  dynamics, star-formation
(SF)  processes, stellar  evolution, and black-hole physics.  A
model that  will combine all the  above physics into a well-defined,
coherent  methodology should thus include a large dynamical range of
scales, from the formation of  individual stars to the formation of
the largest clusters of galaxies.

Historically, a  detailed understanding  of the formation  of
galaxies has  grown   alongside  with  the  development  of   the
concepts  of dark-matter clustering
\citep[e.g.][]{Rees77,Blumenthal84,White91}. It became clear
that galaxy  formation is  a `two-stage  process', where  first
dark matter clumps collapse and virialize into `haloes'. This is
then followed by a  second stage  of  gas cooling into  the central
regions of  these haloes, which can condense and  form stars
\citep{White78}.  In
the last few decades there   have  been substantial   improvements
in   our  quantitative understanding  of structure  formation of
dark-matter, both  from the analytical point of view
\citep{Press74,Bond91,Sheth02} and also using numerical simulations
\citep[e.g.][]{Lacey94,Navarro97,Springel05}.  The current level of
accuracy  in modeling  dark-matter  structure growth  has reached
the level of few percents. Some larger uncertainties are still
inherent in the  halo definition  and  its virial mass,  but
the way these uncertainties affect galaxies is
partially related to the virialization of gas inside haloes, and
thus cannot be fully understood using dark-matter only.

Unlike  the physics  of  dark matter, the  processes  that govern
gas dynamics and SF still remain highly uncertain, usually at the
level of at least an order of magnitude. For example, cooling rates may be
dramatically affected by the assumed shape
of the density profiles or  different assumptions made in computing the cooling function
\citep[e.g.][]{Efstathiou92,Landi99,Maio07,Gnat07, Kaufmann09, Wiersma09}. Also, both
star formation and supernova feedback are very complex processes
whose basic characteristics are still under debate
\citep[e.g][]{McKee77,Dekel86,Elmegreen97,
Maclow04,Murray05,Scannapieco06,Leroy08,Fumagalli09}.

In  view of  the  above  difficulties to  model  galaxies, three
main methodologies were developed in  the literature. The first
approach is a statistical one, attempting
 to interpret  the data with a minimum set of
assumptions.  This includes  variants of  the halo  model, where the
halo and galaxy abundances are linked \citep[see
e.g.][]{Jing98,Peacock00,Seljak00,Cooray02,Kravtsov04,Zehavi05,vdBosch07}.
The approach has
been useful in  quantifying the relationships between  galaxies and
haloes, the clustering  properties of both populations, and  the
properties of satellite galaxies.  However, by definition, this formalism
does not allow to
follow the evolution of  galaxies  with  time,  and  to  study their
formation  histories \citep[but see][]{Drory08,Conroy09}.

A totally different approach is the detailed  modeling  of
individual galaxies  using  high  resolution
simulations which include  gas   hydrodynamics,   and  a   simple prescriptions
for SF, feedback and black hole growth \citep[e.g.][]{Abadi03,Scannapieco09}.  This approach  is  valuable
for testing specific  ingredients of  the models (i.e.  different
feedback mechanisms)   and   it   is   critical  to   deepen our
physical understanding of these processes. However, it is  still difficult
to obtain a statistical sample of galaxies using such
simulations, or to span a large range of
possible models \citep[see a recent attempt by][]{Schaye09}. In
addition,  processes like  star formation and supernova (SN)
feedback occur far below the resolution limits of these simulations,
which makes  it necessary to describe them with analytical and
approximate recipes. The finite  number  of particles representing
the baryonic content of galaxies, and the presence of numerical
artifacts make it difficult to find appropriate recipes which lead
to numerical convergence while producing realistic model galaxies.

The  most widely  used methodology  to interpret  observations  and
to provide  a coherent  picture for  the  physics of  galaxy
formation  is implemented by  Semi-Analytical Models (hereafter
SAMs). The approach was  introduced  by  \citet{White91}  and then
implemented  by \citet{Cole91,Kauffmann93b}.    More
recent    works    include \citet{Somerville99b,Cole00,Benson02,Hatton03,Bower06,Cattaneo06,
Croton06,DeLucia07,Monaco07,Kang05,Somerville08}. These models
provide a detailed picture of galaxy formation processes and produce
a population of simulated galaxies whose
global statistical properties   resemble   the    observations
in   many aspects. Consequently, the models can then be used to
examine specific issues in the formation of  galaxies, like
the impact of different processes on the
shape of the luminosity function  \citep{Benson03},   the  formation
history   of  elliptical galaxies    \citep{DeLucia06},      and
galaxy      merger-rates \citep{Guo08}. SAMs are also valuable for
developing new observational techniques \citep{Chen09}.

The SAM approach is in an intermediate position between the
detailed hydrodynamical simulations and variants of the halo model
discussed above. While in contrast to hydrodynamical simulations, SAMs
include no 3D information on individual galaxies, they follow
the evolution of galaxies within their dark-matter
haloes, which cannot be done using the `halo model'.
In SAMs, halo merger-trees  are  used  as  a skeleton for
the  evolution  of galaxies. The first generation of models used
Monte-Carlo realizations based on the extended Press-Schechter
approach, while modern SAMs use trees extracted from $N$-body simulations.
Recent versions of these models include merger-trees which follow the evolution of substructure
inside haloes. Within these  haloes
galaxies are assumed to grow,  with each  galaxy being described by
a small number  of parameters. This means that only a few mass
components are followed with time: hot gas, cold gas, stars,
black-hole mass, gas which was ejected out  of the  halo, heavy
elements and the fraction of mass contained in the bulge.
SAMs include recipes that are used to compute the rate
of change in each component, according to the properties
of the galaxy and its host halo. For example, the SF rate
is usually assumed to be proportional to the mass of cold gas in the
disk, divided by the disk time-scale.
These recipes then enable  a fast and detailed evolution  of
galaxies within the model. The benefit of SAMs over other approaches
is that using a few simple assumptions  the model can reproduce the
detailed formation histories of galaxies.

Although SAMs have been  quite successful in generating model
galaxies that resemble  observations, there are a few limitations to
this method that we would like to  address.
First, it seems that the degeneracy inherent in these models
has not been sufficiently explored. Often a fixed functional
form is used to describe a given process, and tuning is only
done by varying parameters. In addition, it is not always tested
whether the more complex models can be replaced by simple recipes.
 For example, the physics of AGN feedback is poorly understood, but nevertheless
implemented with a relatively detailed recipe in certain SAMs
\citep[e.g.][]{Croton06,Monaco07}. \citet{Cattaneo06} and \citet{Bower06} have
shown that simply suppressing cooling within massive haloes gives
results which are very similar to the complex models mentioned above
\citep[see however the discussion in][]{DeLucia06}. This means that the
recipe for AGN feedback can be easily summarized by a range of halo
mass and redshift for which the suppression is active. This fact
cannot directly be seen from the detailed equations being used, and
the effective range  in halo mass  where this  suppression  is
active is not  well examined.

A different limitation of SAMs is  that some recipes are hard-wired
into a complex simulation code and inter-connected in non-trivial
ways. It can thus be difficult for users to change the functional
form of these recipes. For example,
cooling efficiencies are computed using information on the density
profile of hot gas within haloes, combined with the physics of
radiative cooling processes. There is no practical
way to change the effective cooling efficiencies for ranges in halo mass and
redshift, and to test its effect on the population of galaxies.
In addition,
testing a large number of very different recipes is nearly
impossible, and it can be very hard or impossible to tune a SAM in
order to match observations if no freedom in the functional forms
is allowed. For example, as  was shown by \citet{Fontanot09},
various  SAMs are unable to reproduce the decrease of  the specific
SF  rate as a  function of stellar mass. Up to now, it has been
unclear if this limitation is fundamental to
the SAM approach  or not. We will show below  that once the
functional shape of the  recipes is relaxed, this  observational
feature can easily be reproduced.

Another  minor  disadvantage is that SAM  recipes depend
occasionally  on  a  large chain  of computational  steps, which  do
not  allow  the reader  to extract  the actual  values being used. For
example,  it is difficult to estimate the actual cooling and SF
efficiencies used by a given SAM, which makes it nearly impossible
to compare different SAMs easily.

In this work we develop a new approach for following a galaxy within
a given merger-tree.  We present   a formalism that  is more
simple  and transparent  than the  usual  implementation of  a set
of recipes, but on the other hand  general and flexible enough so
that  model  galaxies  will  reproduce the observational constraints
accurately. We will show that it is possible to simplify the SAM
considerably by only allowing the recipes to depend on halo mass and
redshift. We find that this  simplification does not reduce the  level
of complexity in the  modeled galaxies, but it enables a very simple
and transparent description of the various physical processes. As a
second step, in  order to keep the  model general and flexible
enough, we do not restrict  the functional dependence of each
process on  the halo mass  and redshift. This generalization greatly
facilitates the tuning process. The formalism we  propose is more
compact  than usual  SAMs  and can be represented  by a  few simple
functions.

The main assumption in our formalism is that each physical process
can be described well  enough by its dependence on the  host halo
mass and redshift.  Few SAM recipes  are already following this
assumption. For example, the SN feedback, including reheating,
ejection and reincorporation of disk gas, only depends on the virial
velocity and virial mass of the host halo in \citet{DeLucia07}.
There are  other recipes which  are  in
general not a  simple function of  halo mass and time. For example,
the cooling  efficiency is
usually modeled  in SAMs by following  the density profile  of the
hot gas component, and  computing the radius within which  gas has had
time to cool. By extracting cooling efficiencies and star formation
efficiencies directly from the SAM of \citet[][hereafter DLB07]{DeLucia07}
we test our main
assumption and show that it is relatively accurate for all the
recipes used in this model.

In this paper we use our  formalism to explore the level of
degeneracy inherent in current galaxy formation models.  We try to examine
very different models  of   galaxy  formation  that  are  all
consistent  with  the observational data.   We deliberately consider
also  some very extreme and physically less well motivated models in
order  to span a large  range of solutions  allowed by
the observations.   This will help us better to understand the
degeneracies involved in SAMs.  Specific questions  can then be
addressed by comparing  the  different  modeled  galaxies.
For example, we derive the minimum cooling efficiencies
which are consistent with the observed population of galaxies;
we examine how much SN feedback is
required in a model with instantaneous cold accretion at low halo
masses; and we investigate a model in which star formation and cooling
terminates after major mergers.

This paper is organized  as follows.
We start with a review of the current
SAM ingredients and their uncertainties in section \ref{sec:motivation}.
In section \ref{sec:formalism} we present  our  general formalism  and
explain  where  it differs from previous works. Our approach is then
compared with a state-of-the-art SAM in \S\ref{sec:test}.
Section \ref{sec:models} summarizes the various ingredients of each specific model we
develop here. These models are further discussed in \S\ref{sec:results} where
their results are compared to observational data. We discuss our results and summarize them  in
\S\ref{sec:discuss}. Throughout this paper  we use the cosmological
model adopted  by the Millennium simulation  with
$(\omm,\,\oml,\,h,\,\sigma_8) =(0.25,\,0.75,\,0.73,\,0.9)$.  We  use
$\log$ to designated $\log_{10}$, $t$ is the time in Gyr since the big-bang.


\section{Motivation: Uncertainties in SAM ingredients}
\label{sec:motivation}

In the models presented  in this work we change both the absolute values and the
parameterization of the efficiencies of star formation, SN
feedback and cooling more freely than what is usually done in standard semi-analytical models. We
also allow  the  prescriptions for dynamical  friction and
merger-induced  star-bursts to vary in  part of our  models.  Here we  give a
brief explanation why we believe  the standard recipes might be overly
restrictive. Although these recipes may also be modified
within standard SAMs, we argue that
our formalism allows easier and better controlable changes.

\subsection{Cooling}
\label{sec:cooling}

The  efficiency with which the hot gas radiates and cools
is  normally not  subject to any  tuning in  SAMs.
Cooling efficiencies are  calculated according  to \citet{White91},
using  cooling functions from  \citet{Sutherland93} and  assuming  an
isothermal  or NFW gas  density  profile\footnote{
Note however,  that  some
freedom  is left  for fixing the age of the halo
which determines the radius inside which gas is allowed to cool.
The age of the halo  might
corresponds to the age of the universe, the last major merger event,
the first time the halo was identified, or the dynamical time of the halo.}.
This makes cooling  very efficient,
which is the reason that strong SN and AGN feedback are needed
in all SAMs. Next  to the  dark  matter  physics,
cooling  rates  are  probably  the most  fundamental  ingredient
in SAMs  and shape many of  the secondary ingredients.
However, various complications  challenge the generally used cooling
prescription.

The assumed density profile of the hot gas within haloes can
greatly affect the cooling efficiencies,
as cooling depends strongly on the local gas density.
Only a few semi-analytical models differ from
the standard prescription by assuming a different gas density profile
\citep[e.g.][]{Cole00} or considering the dynamical adjustment of the
density profile \citep{Cattaneo06}.
However, as shown by \citet{McCarthy08b} and \citet{Kaufmann09},
assuming  a cored  gas density  profile instead  of the  commonly used
steeper profiles greatly decreases cooling rates,
and thus the need for strong  AGN and SN feedback both in clusters and
Milky-Way  sized haloes.   Various `preheating'  mechanisms  which may
decrease  the  central  density  and  entropy of  the  gas  accordingly
have  been
suggested.  These include: gravitational pancaking \citep{Mo05}; early
quasar  feedback \citep{Lu07,  McCarthy08b}; feedback  produced  by an
early star burst \citep{Tang09};  and the early chaotic assembly phase
of haloes \citep{Keres09}.

At gas  densities
typical  of the  hot gas  haloes  around low mass
galaxies,  the assumption  of
collisional ionization equilibrium that is used to calculate the
\citet{Sutherland93} cooling functions may break down
\citep{Gnat07,Wiersma09}. This may decrease cooling rates
by  roughly an  order  of magnitude. The  effect could  even be  more
dramatic in the  presence of radiation emanating from  an AGN or young
stars residing  in the  galaxy \citep[see the SAM by][]{Benson02}.
Departing from non-solar abundance
ratios can also affect cooling rates by a factor of a few \citep{Smith08,Wiersma09}.

Finally, as SAMs only assume spherical density gas profiles, the rich
three-dimensional effects induced by e.g. cold filaments cannot be addressed
in detail. Several comparisons were made between hydrodynamical simulations and
the simplified SAM approach, mostly finding a reasonable agreement
\citep{Benson01,Yoshida02,Helly03,Cattaneo07}.  On the other hand,
 \citet{Viola08} claim that details in the calculation might
change cooling rates by an order of magnitude at high redshift.

\subsection{Star formation}

Most SAMs assume that the  efficiency with which cold gas is converted
into  stars is  inversely proportional  to  the dynamical  time of  the
disk. In some  models a critical threshold density of  the cold gas is
used, below  which stars are not  able to form.  These  laws replace a
large  amount of  complicated physics which cannot be traced in detail
within SAMs. Modern  observational studies  of  SF in  local
galaxies  stress  that  different  components  may  play  a  role  in
regulating  SF  within disks:  the  atomic  gas  (\HI), molecular  gas
(H$_2$,  found inside  giant molecular  clouds), and stars,
which are all interconnected \citep[e.g.][]{Wong02,Fumagalli09}.

Empirical laws may  be derived from low redshift  observations and can
then  be  used  by  SAMs  to  model  galaxies  at  high-$\z$  as  well.
SAMs generally make use of the empirical Schmidt-Kennicutt law
\citep{Kennicutt98a, Schmidt59}, which relates the star formation rate
to the total cold gas surface density, i.e. the \emph{sum} of the \HI\ and H$_2$ masses.
However, the uncertainty   inherent  in  this law  is  large  \citep{Boissier07}.
Both \citet{Wong02} and \citet{Leroy08} argue in favour of a Schmidt law based on the surface
density of the molecular gas alone.
\citet{Leroy08} found  that in relation to the
total gas density, SF  efficiencies can vary between $10^{-2}$ up to
1 Gyr$^{-1}$.  Modeling the two  phases of cold gas seems difficult as
it involves  pressure estimates  within the disk  \citep{Blitz06}, and
the  formation physics  of  giant molecular  clouds (but see
\citet{Obreschkow09} and \citet{Dutton09} for recent attempts).
Lastly, the existence of a threshold in gas density
for SF  that was suggested by \citet{Kennicutt98a}  is still debated
\citep{Schaye04,Boissier07}. In particular, including such a threshold might
not be necessary if SF is assumed to depend on the molecular gas density instead of
the total gas density, since the molecular fraction is suppressed at low densities \citep{Dutton09}.

As mentioned above, observations find correlations between SF rates and the cold gas
surface density and not the total mass of cold gas in the disk.
This is modeled in most SAMs by computing the disk radius
and disk dynamical time using the halo spin parameter,
and assuming a smooth gas distribution.
We do not follow this
method here, which essentially means that we assume some average disk radius
in a fixed bin of halo mass and time.
This seems to be a reasonable assumption as the spin parameter shows a
small dependence on halo mass, along with a large general scatter \citep{Bett07}.
This means that not considering the spin parameter mainly reduce the scatter
in the SF efficiency. We will show below that this
does not change the properties of the model galaxies significantly.

\subsection{SN feedback}

Some sort of feedback by stellar processes is an ingredient in virtually all
semi-analytical and  hydrodynamical models of galaxy  formation.
Suggested feedback processes include heating by SN or photoionization
radiation of massive stars, momentum deposition by SN, stellar
winds, expanding bubbles of photoionized H\textsc{ii} regions, and
absorption and scattering of starlight by dust grain \citep[see the discussion in][and
references therein]{Murray09}.
Most SAMs only attempt to model the feedback by SN,
and its detailed implementation differs greatly between different
models and is not well constrained by observations
\citep[see e.g. the discussion in][]{DeLucia04}.
Many SAMs combine  two different modes of
SN feedback: reheating, in which  cold disk gas is transferred back into
the hot mode with a rate equal to a few times the star formation rate;
and ejection,  in which part  of the hot or cold
gas is assumed to  leave the
halo, where it only becomes  available again for cooling after roughly
one  dynamical time \citep[e.g.][]{DeLucia04,Croton06, Bower06, Somerville08}.
Other SAMs do not include any ejection
\citep[e.g.][]{Kauffmann93b}, or assume that part of the ejected gas
can never come back to the hot phase \citep{Somerville99b,Bertone07}.

The efficiency  of reheating and ejection adopted in SAMs is mainly
determined by  the need to reconcile  high cooling rates  with the low
number  of  stars  observed.   It  is only  vaguely  based  on  direct
observational  constraints of  outflows in  galaxies. For  example, the
reheating efficiencies used by DLB07  are based on the observations by
\citet{Martin99} of  strongly star-bursting  systems, which  are
a population of galaxies with properties at
the extreme of the distribution. However, there
are indications that  outflows can
only  be generated  above some  limiting star  formation  density
\citep[e.g.][]{Strickland09} and that it is thus not appropriate to
apply the same recipe to all galaxies. Note also that stellar
feedback by other sources than SN might scale very differently with
galaxy properties.

\subsection{Merger-induced SF bursts}

The efficiency of SF in merger induced bursts is defined as the fraction of
cold gas converted into stars during the merger event, after subtracting the amount
of stars produced in the quiescent mode. Numerical simulations which include
SF, feedback and cooling can be used to quantify the SF efficiency
\citep{Cox06,Cox08}. Although the efficiency computed by these simulations
usually agrees with the standard recipes used in SAMs (Eq.~\ref{eq:sf_burst}),
the dependence on additional parameters like the shape of the orbit and bulge
mass might propagate systematic errors into the SAMs. In addition, the galaxies
being simulated are generally assumed to be typical low-redshift
galaxies. Efficiencies at high $\z$ are still poorly constrained.

Numerical simulations indicate that the time-scales for burst duration depend
strongly on the feedback recipe being used. For example, \citet{Cox08} examine
two different feedback recipes which deviate by a factor of $\sim 2-5$ in the burst
time-scale. In general, the burst duration in these simulations can vary between 0.1 and 2 Gyr,
and its behaviour at high-$\z$ has yet to be investigated.

In most current SAMs, the SF burst  is
assumed  to start \emph{after}  the galaxies  have coalesced  into one
remnant  galaxy. This  is  very different  from numerical  simulations
where  bursts can be  triggered  at intermediate
close passages between galaxies. This effect may limit the SAM's ability
to   use   proper  time-scales   for   bursts  at the accurate time
within the merger event. In order to use long time-scales which
mimic few several
 small bursts, one may need to  decrease the dynamical-friction time,
so that the burst timing will be accurate.

\subsection{Dynamical friction timescales}
\label{sec:df_uncertainties}

In  order to  properly model  the behaviour of
satellite galaxies, all SAMs need an analytical   prescription
for dynamical friction. If merger trees from
$N$-body simulation are used, such a prescription is necessary because
subhaloes may fall below the resolution limit due to tidal
stripping, which means their orbits cannot be traced
anymore. If \fof merger trees are used, on the other hand,
it is needed as soon as a galaxy becomes a satellite since no orbital information
is available thereafter.

The time  it takes galaxies to  sink into the center of
the  host potential well  and then  merge with the central galaxy  is
usually estimated by the Chandrasekhar formula (see Eq.~\ref{eq:t_df}).
Recent studies are  able  to   estimate the timescale more
accurately    using    high    resolution    simulations
and taking into account detailed orbital
information \citep{Boylan08,Jiang08}. However,  the  orbital
information is  usually not considered by SAMs.  An additional problem
arises  because
it is unclear how the mass of the sinking galaxy/subhalo should
be estimated; using the mass of the subhalo when it was last
identified or the baryonic mass of the galaxy instead
changes the dynamical friction timescales significantly.
This can be seen from Eq.~\ref{eq:t_df} below, where the dependence
on the satellite mass is strong.

These  problems may bias the estimates  of the dynamical
friction timescale  high or low.  Indeed, DLB07 had to  increase their
dynamical  friction   timescales  by  a   factor  of  2   compared  to
\citet{Croton06} in order to fit the observational data. It may well be that other models of galaxy formation
would  require  a  higher  correction  factor,  or  one  that  depends
non-trivially on  time or halo  mass.

\subsection{Environmental Effects}

In most SAMs galaxies lose their entire
hot gas reservoir as soon as they become satellites inside a \fof group,
and this hot gas then becomes directly available for cooling to
central galaxies. This leads to satellites which are too red compared to observations
\citep[e.g][]{Weinmann06a, Baldry06, Kimm09} and may also lead to an overestimate
of hot gas that is available to cool to the central galaxy. An improved SAM which is more
successful in reproducing observations in this respect has been presented
by \citet{Font08}.

Environmental effects are subject to large uncertainties, as
efficiencies of ram-pressure and tidal stripping are not well known
and depend on the detailed orbit of the satellite {\citep{McCarthy08}.
Environmental effects
are also strongly linked to the detailed prescription for SN feedback in
satellite galaxies \citep[e.g.][]{Okamoto09} and the resulting
gas density profiles of the satellites. Furthermore, it is unclear
if stripped satellite gas should be made available for cooling to the central
galaxy in all cases; a large fraction of galaxies commonly
identified as satellites in \fof groups reside outside the virial
radius of the central galaxy which makes it unlikely that their gas
reservoir is immediately added to the gas reservoir of the central galaxy. One promising
approach seems to be to strip the diffuse gas around satellites in
proportion to the dark matter \citep{Weinmann09}.
Here, we use a more simplified method and strip the hot gas around satellites
exponentially, with an exponential timescale of 4 Gyr.
Note that we do not include a prescription for the disruption
of the cold gas or stellar component of the galaxy,
which might be important for reproducing the correlation function on
small scales or halo occupations statistics
\citep[e.g.][]{Henriques08, Kim09}. Such a recipe could however
easily be incorporated into our methodology.


\section{The Formalism}
\label{sec:formalism}

In this section we describe in detail the formalism we develop for
modeling the formation and evolution of galaxies. This formalism is
very general and is not restricted to a given model\footnote{We use
the term `formalism' to describe the general methodology we use,
specific solutions that can be applied within this framework are
called `models'}. Consequently, we do not describe in this section
the functional shape of each process, or the parameters being used.
The motivation in setting up this approach is to have a methodology
which is conceptually simple, includes most of the physical
processes, and is general enough so that very different models can
be described within the same language.

\subsection{Merger trees}

We use merger trees extracted from the Millennium
$N$-body simulation \citep{Springel05}. This simulation was run
using the cosmological parameters
$(\omm,\,\oml,\,h,\,\sigma_8)=(0.25,\,0.75,\,0.73,\,0.9)$, with a
particle mass of $8.6\times10^8\,\hmsun$ and a box size of 500
$h^{-1}$Mpc.
The merger trees used here are based on \emph{subhaloes} identified using the
\textsc{subfind} algorithm \citep{Springel01}. They are defined as the bound
density peaks inside \fof groups \citep{Davis85}.
More details on the simulation and the subhalo
merger-trees can be found in \citet{Springel05} and
\citet{Croton06}. The mass of each subhalo (referred to as $M_h$ in what follows)
is determined according to the number of particles it contains.
Within each \fof group the most massive subhalo  is termed
the central subhalo of this group.
Throughout this paper we will use the term `haloes' for both subhaloes and
the central (sub)halo of \fof groups.
In general, our approach could also be based on \fof merger-trees, which
might be generated by Monte-Carlo realizations.

\subsection{Quiescent evolution}

Each galaxy is modeled by a 3-component vector,
\begin{eqnarray}
\m = \left( \begin{array}{c} \ms \\ \mc \\ \mh \end{array} \right)
\, ,
\label{eq:m_vec}
\end{eqnarray}
where $\ms$ is the mass of stars, $\mc$ is the mass of cold gas, and
$\mh$ is the mass of hot gas distributed within the host halo. We
use the term `quiescent evolution' to mark all the evolutionary
processes of a galaxy, except those related to mergers.
The quiescent contribution due to each physical process to the components of
$\dot{\m}$ is designated by square brackets. For example,
$\left[\dotmc\right]_{\rm feedback}$ is the change
in the mass of the cold gas due to feedback.

Fresh supply of gas into the galaxy is provided by smooth accretion
along with the growth of the dark-matter halo mass. The rate of
cold and hot accreted gas is modeled by
\begin{eqnarray}
\left[\dotmc\right]_{\rm accretion} &=& \fca\cdot \dot{M}_h \\
\left[\dotmh\right]_{\rm accretion} &=& \fha\cdot \dot{M}_h \, .
\label{eq:smooth_acc}
\end{eqnarray}
Here $\dot{M}_h$ is the rate of dark-matter smooth accretion
which does not include mergers with resolved progenitors (if $\dot{M}_h<0$ we use a
gas accretion rate of zero). The fractions $\fca$ and $\fha$ are assumed to be
some functions of the halo mass and redshift, with the standard
values being 0 and 0.17 respectively (note that the sum $\fca+\fha$ corresponds to
the cosmic baryon fraction in all the models presented
in this paper).

The mass of cold gas may increase due to radiative cooling of the
hot gas. This cooling rate is modeled by
{\begin{equation}
\left[\dotmc\right]_{\rm cooling} =  -\left[\dotmh\right]_{\rm cooling} = \fc \cdot \mh \,.
\end{equation}
The cooling efficiency, $\fc=\fc(M_h,t)$, is a function of the host
halo mass $M_h$ and the cosmic time $t$ only, and given in units of
Gyr$^{-1}$. This is a
significant simplification over other approaches, neglecting the
additional dependence on the gas metallicity and density, which
is only implicitly included within the halo mass dependence.
We will examine the difference between our approach and the
usual SAM recipe in section \ref{sec:test} below.

We assume that the SF rate is proportional to the amount of cold gas,
\begin{equation}
\left[\dotms\right]_{\rm SF} = -\left[\dotmc\right]_{\rm SF} = \fs\cdot \mc \,,
\label{eq:sf}
\end{equation}
where $\fs=\fs(M_h,t)$ is a function of the halo mass and time, in units of Gyr$^{-1}$.
For each SF episode we assume that a constant fraction of the mass is returned
back to the cold gas component due to SN events and stellar
winds. This recycling is assumed to be instantaneous,
and contributes
\begin{equation}
\left[\dotmc\right]_{\rm recycling} = -\left[\dotms\right]_{\rm recycling} =
R \left[\dotms\right]_{\rm SF} \,.
\end{equation}
Here $R$ is a constant with typical values in the range $0.4-0.7$.
Stellar population synthesis models
\citep[e.g.][]{Bruzual03} predict a recycling time of approximately
1 Gyr. As a result, our assumption of instantaneous recycling might
change the cold gas fractions inside galaxies, and consequently the
values of SF rates. We plan to test this assumption in a future
work.

Gas can be heated due to SN explosions and move
from the cold gas into the hot. Assuming that SN events
immediately follow star formation, this feedback should be in
proportion to the SF rate,
\begin{eqnarray}
\lefteqn{
\left[\dotmh\right]_{\rm feedback} =  -\left[\dotmc\right]_{\rm feedback} = } \\ \nonumber & &
\ffd\left[\dotms\right]_{\rm SF} = \ffd \fs \mc \,.
\end{eqnarray}
Here feedback is modeled by a function of halo mass and time $\ffd=\ffd(M_h,t)$}.
Other feedback mechanisms like feedback by AGN are
not in general proportional to the SF rates, and are not
explicitely modelled in our approach (although may be regarded
as implicitely included in the net cooling rates). Note that we assume
feedback is proportional to the value of SF
rate before recycling is considered. We allow some of the gas to be ejected from the hot
gas phase out of the halo:
\begin{equation}
\left[\dotmh\right]_{\rm ejection} =  -\fe \left[\dotms\right]_{\rm SF} = -\fe \fs \mc \,,
\end{equation}
where again the efficiency $\fe$ is a function of halo mass and time only.
This gas is never allowed to fall back, so the baryonic mass inside haloes
might be lower than the universal average.

To conclude, each process is described by one function which depends
on the host halo mass and time only. All processes
discussed in this section can be
written in a compact form by using the following differential equations:
\begin{equation}
\dot{\m} = \A\m + \B\dot{M}_h \, ,
\label{eq:m_evolve}
\end{equation}
where
\begin{eqnarray}
\A = \left( \begin{array}{ccc}
0 & (1-R)\fs & 0 \\
0 & -(1-R)\fs -\ffd \fs & \fc \\
0 & \ffd \fs - \fe \fs & -\fc
\end{array} \right)
\end{eqnarray}
\begin{eqnarray}
\B = \left( \begin{array}{c}
0  \\
\fca  \\
\fha
\end{array} \right)  \,\, .
\label{eq:AB_defs}
\end{eqnarray}
We emphasize that $\A$ and $\B$ do not depend on $\m$, resulting in
a set of linear inhomogeneous differential equations.

Photoionization heating of the intergalactic medium is assumed to
suppress the amount of cold gas available for SF within low mass
haloes. This effect is critical for modeling the formation of dwarf
galaxies. The minimum halo mass of $\sim 2\times 10^{10}\,\hmsun$ in
the Millennium simulation, which we use here, does not allow a
detailed modeling of small mass galaxies. Consequently, we use the
simple assumption that all the gas is kept hot until redshift 7
when cooling and SF processes can start. This information is
included in the components of $\A$ and $\B$.

\subsection{Mergers and satellite galaxies}
\label{sec:formalism_mergers}

Satellite galaxies are defined as all galaxies inside a
a \fof group except the main galaxy inside
the central (most massive) (sub)halo. Once the (sub)halo corresponding
to a given galaxy cannot be resolved anymore, it is considered
as having merged with the central halo. Due to the effect of dynamical friction,
the galaxy is then assumed to spiral towards the center of the
\fof group and merge with the galaxy in the central halo after a
significant delay time.

At the last time the dark matter (sub)halo of a satellite galaxy
is  resolved we compute its distance from the
central halo ($r_{\rm sat}$), and estimate the dynamical friction
time using the formula of \citet{Binney87},
\begin{equation}
t_{\rm df} = \alpha_{\rm df} \cdot \, \frac{1.17 V_v r_{\rm sat}^2}{G m_{\rm sat}\ln\left(
1+ M_h/m_{\rm sat} \right) } \, .
\label{eq:t_df}
\end{equation}
For $m_{\rm sat}$ we use the baryonic (stars + cold gas) mass of the
satellite galaxy plus the minimum subhalo mass which can be resolved by the
Millennium simulation.
$V_v,\,M_h$ are the virial velocity and mass of the central subhalo.
We add a free parameter $\alpha_{\rm df}$ to this equation in order to reflect
the uncertainties discussed in section \ref{sec:df_uncertainties}.
If a satellite falls into a larger halo together with its central galaxy we update
$t_{\rm df}$ for both objects according to the new central galaxy.

Given the simplifications used in the previous sections, one
might argue that $t_{\rm df}$ should be formulated a function of halo mass
and redshift only as well, without assuming any a-priori parameterization.
However, it is very likely
that such a function should also depend on the satellite mass,
$m_{\rm sat}$, resulting in at least three variables for this function. This
variable space might be too large for finding useful constraints.
In addition, the level of accuracy in dynamical friction
estimates is becoming high enough, and will leave less room for freedom in the near future.

While satellite galaxies move within their \fof group, they suffer
from mass loss due to tidal stripping. The hot gas halo of the
satellite should be the first baryonic component that will be
stripped. We assume that all satellite galaxies
are losing their reservoir of hot gas exponentially, on a time scale of a few Gyr.
In order to properly model this stripping we modify $\A$
by subtracting a constant $\alpha_h$ from one of its elements:
\begin{eqnarray}
\A_{sat}(3,3) = -\fc-\alpha_h \,.
\label{eq:ABsat_defs}
\end{eqnarray}
Note that a constant in the diagonal of $\A$ gives an exponential
time dependence. However, the actual
dependence of $\mh$ on time for satellite galaxies is more
complicated due to contributions from feedback and cooling.
In general the parameter $\alpha_h$ should depend on the dynamical
time of the host halo. For simplicity we consider it to be a constant here.

When galaxies finally merge we assume that a SF burst is
triggered. We follow \citet{Mihos94,Somerville01,Cox08} and
model the amount of stars produced by
\begin{equation}
\Delta \ms = \alpha_b \left( \frac{m_1}{m_2} \right)^{\alpha_c} (m_{1,{\rm cold}}+m_{2,{\rm cold}}) \, .
\label{eq:sf_burst}
\end{equation}
Here $m_i$ are the baryonic masses of the progenitor galaxies (cold
gas plus stars), $m_{i,{\rm cold}}$ is their cold gas mass, and
$\alpha_b$, $\alpha_c$ are constants.

The efficiency is computed
\emph{before} applying the recycling factor $R$, so the net amount
of stars added to the galaxy is $\Delta \ms(1-R)$. SF in bursts are
accompanied by the same SN feedback as used for the quiescent
mode of SF. The burst duration $\sigma_{\rm burst}$, may vary between tens of Myr to
a few Gyr depending on the merger mass ratio, and whether
multiple bursts are considered or just the main peak  \citep{Cox08}.
We will specify below the time-scale adopted for each model.

\subsection{Usage: degrees of freedom, tuning, code}
\label{sec:usage}

The number of degrees of freedom available for tuning is much larger
here than in any other SAM which assumes some fixed functional form for the recipes used.
The quiescent evolution part includes six functions,
$\fc,\;\ffd,\;\fs,\;\fe,\;\fca,\;\fha$ where each may have any
dependence on halo mass and time. There are additional few constants:
$R,\;\alpha_{\rm df},\;\alpha_h,\;\alpha_b,\;\alpha_c$ which are mostly
related to mergers and satellite treatment. The large freedom inherent in
the formalism allows us to incorporate very different models, but it demands
some starting point when tuning the model against observations. As will be demonstrated
below, for each model developed here most of the free parameters are fixed
a-priori and are not tuned. Once a set of such assumptions is made we can explore
the freedom in one or two components, and try to match the observational data.

Our code is available for public usage through the internet (see
\texttt{http://www.mpa-garching.mpg.de/galform/sesam}). Users are allowed
to use any functional shapes for the quiescent ingredients by loading
tables of efficiency values as a function of time and halo mass.
Alternatively, simple power-law functions can be used.
The current public version applies the model to a small simulation volume\footnote{
We use the Milli-Millennium
simulation, which was run with the same parameters as the full
Millennium simulation, but within a box size of 62.5 Mpc $h^{-1}$
\citep{Lemson06}.}
and provides results in just a few seconds. It also automatically compares
the model results to observation.

In order to solve Eqs.~\ref{eq:m_evolve} -- \ref{eq:AB_defs} we
divide the time-step between each simulation snapshot (which is
roughly 250 Myr in the Millennium simulation used here) into 20
smaller steps ($\sim12$ Myr each). In each such small step we compute
the incremental addition to the mass of stars, cold gas, and hot
gas. We checked that using different number of time-divisions
(5-100) does not change the results of our models
significantly. This means that at any time-step the change in the masses of the galaxy
components is small, and the models are numerically converged.
An error message will appear in case the recipes for quiescent
evolution produce negative masses. In this case, either the
time-step, which is a tunable parameter, has to be increased, or
the recipes have to be adapted.
We found that convergence is achieved once the efficiencies values
are smaller than one over the time-step size.

Unlike in the quiescent evolution, negative masses are occasionally produced by our recipes
describing star-bursts. Our models usually assume that these bursts
occur rapidly on time-scales of few tens of Myr. As a result, the
feedback efficiency might be so high
that it would be able to heat up more cold gas than is actually
present within the galaxy. In these cases we simply limit
the amount of heated gas to the amount of cold gas available after
producing the stars in the burst.

\section{Comparison to a full SAM}
\label{sec:test}

In this section we compare our formalism to a full semi-analytical
model (SAM). We examine how the  results of the SAM change if the
recipes are simplified to depend only on halo mass and redshift. The
resulting change in the properties of the simulated galaxies will
hint at the level of complexity that can be achieved by our
approach. This will also indicate if our formalism can serve as a
common language to describe other SAMs published in the literature.

We choose to analyze as a reference the SAM developed by DLB07. This
makes the comparison simple because the same set of merger-trees are
being used, and because the results of DLB07 are publicly available.
We introduce a model which follows
the formalism introduced in section \ref{sec:formalism}
and resembles DLB07. It will be named `Model 0' hereafter.
Our general approach is different
from the DLB07 model mainly in parts related to the quiescent
evolution. Other components of Model 0 which are related to
mergers and satellite galaxies can be easily matched to resemble the
details of the DLB07 model. We list these minor adjustments below:
\begin{itemize}
 \item    Dynamical   friction:    A   few    modifications   to
Eq.~\ref{eq:t_df}  are  made in  order to  closely follow
DLB07: Within Model 0, $m_{\rm sat}$ is the dark-matter
halo mass of the satellite  galaxy at the  last instance
when it was  still resolved  instead  of  the  baryonic
mass  of  the satellite used in other models in \S\ref{sec:models}
below. We also do not  allow $r_{\rm sat}$ to
be larger than  the virial  radius of  the
central galaxy and we use $\alpha_{\rm df}=2$.
 \item Satellite stripping: we set $\alpha_h$ from
Eq.~\ref{eq:ABsat_defs} to a very large number. This makes
galaxies lose their hot gas to the central galaxy within the
\fof group immediately as they enter the group.
 \item Merger-induced SF bursts: once the galaxies merge, the efficiency
of SF bursts is implemented according to Eq.~\ref{eq:sf_burst}
with $\alpha_b=0.56$ and $\alpha_c=0.7$.
\end{itemize}
We do not modify our implementation of reionization in order to
mimic DLB07 accurately, but keep it as
described in our section \ref{sec:formalism}. We verified that
changing the details of this recipe does not affect the results that
we will show below.

\begin{figure}
\centerline{ \hbox{ \epsfig{file=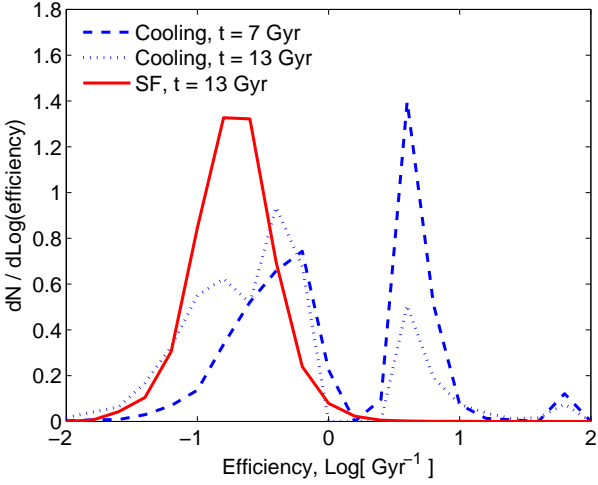,width=9cm} }}
\caption{The full distribution of efficiencies for a given halo mass and time,
as extracted from the DLB07 model.
We plot the SF efficiency ($\fs$) in solid line for galaxies identified at $\z=0$.
$\fc$ is plotted with dotted (dashed) line for $t=13$ (7) Gyr after the big-bang.
All galaxies are central inside their \fof groups, with host subhalo mass
which equals $10^{11}\,\hmsun$.
The distribution of cooling efficiency is bi-modal (see text).}
  \label{fig:eff_dist}
\end{figure}

\subsection{Model 0: ingredients of the quiescent evolution}

A galaxy within the DLB07 model is assumed to have four
main components: hot gas, cold gas, stars, and ejected gas. The
ejected gas is assumed to have been ejected out of the hot gas halo
due to the energy input by SN feedback. It will only be
re-incorporated into the hot gas (and be susceptible to cooling again)
after a certain delay time. The inclusion of such a phase thus
effectively decreases the amount of gas that can cool to
the galaxy. The ejected gas component is not a part of our
formalism as presented in section \ref{sec:formalism}
because it will not be used in the new models we will develop below.
However, we can easily add it here and change the number of
components in the model. This is
implemented by replacing Eq.~\ref{eq:m_vec} with:
\begin{eqnarray}
\m_{D} = \left( \begin{array}{c} \ms \\ \mc \\ \mh \\
m_{\rm ejct} \end{array} \right) \, .
\label{eq:m_vec_DLB07}
\end{eqnarray}
We also need to increase the number of differential equations:
\begin{eqnarray}
\A_D = \left( \begin{array}{cccc}
0 & (1-R)\fs & 0 & 0 \\
0 & -(1-R)\fs -\ffd \fs & \fc & 0 \\
0 & \ffd \fs - \fe \fs & -\fc & \fr \\
0 & \fe \fs & 0 & -\fr
\end{array} \right) \, .
\label{eq:AB_defs_DLB07_A}
\end{eqnarray}
Here $\fr$ is the efficiency for re-incorporation of gas from the ejected
phase back to the hot gas within the halo. We use
\begin{eqnarray}
\B_D = \left( \begin{array}{c}
0  \\
0  \\
0.17  \\
0
\end{array} \right)  \,\, ,
\label{eq:AB_defs_DLB07_B}
\end{eqnarray}
which means that all the fresh gas that is accreted along with the
smooth dark-matter is added to the hot phase first.

Some of the ingredients of $\A_D$ are already given by DLB07 simply
as functions of halo mass and time. These include the feedback
efficiency $\ffd=3.5$; as well as ejection efficiencies ($\fe$ and
$\fr$) which are described by \citet{Croton06}, Eqs.~20 \&
21\footnote{We found that $\fe=27.48 M_{10}^{-0.66} t^{0.65}-3.5$ and
$\fr=2.86 t^{-0.927}$. Here $M_{10}$ is the subhalo mass in units
of $10^{10}\,\hmsun$ and $t$ is in Gyr.}. On
the other hand, the cooling ($\fc$) and SF ($\fs$) efficiencies are
complicated and depend on various extra parameters. In order to
replace these recipes with functions of halo mass and time we run
the DLB07 model over the Milli-Millennium simulation.
We record SF and cooling efficiencies at each
time-step for all galaxies which are central within their
\fof group. This ensures that our cooling is not affected by the
specific treatment of satellite galaxies. Median values are then
computed for bins of halo mass and redshift.
We describe below some details of these recipes, and the values we adopt.

The cooling efficiency in the DLB07 model is computed according to
the growth of the `cooling radius', which describes the maximum
radius at which  the hot gas density is still high enough for the
cooling to occur within a halo dynamical time. This radius is
determined using the efficiency of radiative
cooling as given by \citet{Sutherland93}, which depends also on the
metallicity of the hot gas.
Moreover, as the density profile of the hot gas is assumed to adjust itself
instantaneously to the total amount of hot gas, cooling rates goes like $\fc\mh^{1.5}$ and
not like $\fc\mh$ as assumed here. As a result, cooling efficiencies are
not simply a function of halo mass and time. In addition, AGN
feedback can decrease the cooling efficiency in the DLB07 model, and
practically shuts off all cooling for high mass haloes.

In Fig.~\ref{fig:cool_effic} we plot median values of cooling
efficiencies from the DLB07 model according to the procedure
outlined above. For each galaxy we first average the efficiency
values within one snapshot, going over 20 time-steps which are used
to integrate the model equations. As seen from the figure, gas is
not allowed to cool in haloes with masses above $\sim
10^{12}\,\hmsun$. This is due to AGN feedback implemented by DLB07.
For lower halo masses the cooling efficiency is almost constant with
halo mass, and increases dramatically only at early times ($\z>1$,
$t<7$ Gyr). In order to mimic accurately these cooling efficiencies
we take the specific numerical values shown in
Fig.~\ref{fig:cool_effic} and use them to define $\fc$ in our model.
These values are also given explicitly in Appendix A.
We have checked the dependence of cooling on gas metallicity in
DLB07 and found it to be small for a given halo mass and redshift.

Efficiencies shown in Fig.~\ref{fig:cool_effic} are median values
computed from a large statistical sample of galaxies. It is
interesting to examine the full distribution of $\fc$, for a given
bin of time and mass. In Fig.~\ref{fig:eff_dist} we show the
distribution of cooling efficiencies from DLB07, for haloes of mass
$\sim$ $10^{11}\,\hmsun$. Surprisingly, the cooling efficiency is
bi-modal, showing two distinct peaks, even though the halo mass is
within the `fast cooling' regime, where cooling should be highly
effective \citep[see][]{Birnboim03,Croton06}. We explored this
feature in detail, and found that the bi-modality comes from the
evolution of the cooling radius with time. All galaxies that live
inside small mass haloes experience periods of fast and slow cooling
rates within a few time-steps (the fast cooling episode usually last
one time-step, $\sim$10 Myr). This is due to the fact that the
cooling radius is proportional to the square root of the mass of the
hot gas, {and gas is assumed to be distributed instantaneously within
the halo  \citep[see][Eq.~3 \& 4]{Croton06}.

\begin{figure}
\centerline{ \hbox{ \epsfig{file=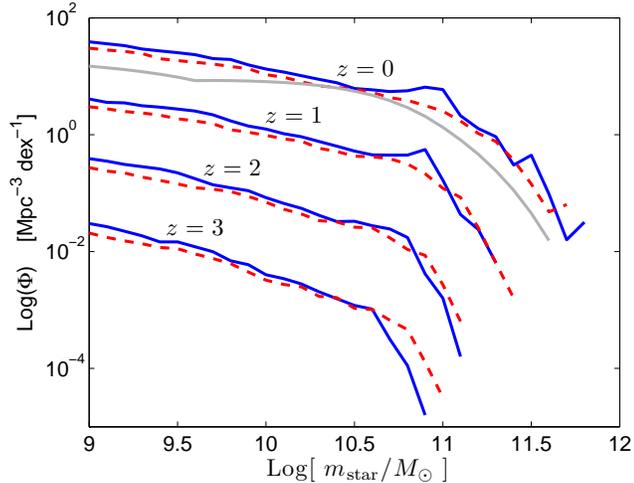,width=9cm} }}
\caption{Stellar mass functions for Model 0 (solid) compared to
DLB07 (dashed). Both samples are based on the Milli-Millennium simulation. The thick
gray line at $\z=0$ shows the observational results of \citet{Li09}.
For each different redshift we multiply the y-axis by a constant for clarity.}
  \label{fig:gsmf_DLB07}
\end{figure}

Both DLB07 and our model assume that the SF is proportional to the mass
of cold gas in the galaxy. However, DLB07 uses an estimate of the
disk radius which depends on the spin parameter of the halo, and cannot be
represented by a simple dependence on halo mass and time. In
addition, DLB07 incorporate a threshold mass of cold gas, below
which stars cannot form. This threshold mass is defined by DLB07
to be a function of halo mass and time only. Consequently, in this
section we will use a modified SF law, namely
\begin{equation}
\,\,\,{\rm SF_{D}:}\,\,\,  \dot{m}_{\rm star} = \fsdb07 \cdot (\mc-m_{\rm crit}) \,.
\end{equation}
Using the above method to record cooling and SF efficiencies
we found that the median SF efficiency used by DLB07 is accurately
described by an analytic fitting function:
\begin{equation}
\fsdb07 = M_h^{1.04}\; t^{-0.82} \; 10^{-6.5 -0.0394 \left[\log M_h\right]^2 }    \,.
\label{eq:fsdb07}
\end{equation}
It should be noted that the two contributions from $M_h$ almost
cancel each other out, which means that
the overall dependence on $M_h$ is small, as can be seen in Fig.~\ref{fig:sf_effic}.
In Fig.~\ref{fig:eff_dist} we show the full distribution of SF
efficiencies for a given bin of mass and time. This distribution is
relatively narrow, and does not show any significant features.

The threshold mass used by DLB07 can be parameterized
as a function of halo mass and time by
\begin{equation}
m_{\rm crit} = \fsdb07^{-1} \; 10^{-8.61}\; M_h^{0.68}\; t^{-0.515}  \,.
\end{equation}
Here $t$ is the time since the big-bang in Gyr, and $M_h$ is the
subhalo mass in units of $\hmsun$. We use a recycling factor,
$R=0.43$, as is done in DLB07.


\subsection{Model 0: simulated galaxies}

Our model is based on a completely different numerical code than the
model of DLB07. In order to perform the comparison below we applied
our Model 0 to the same merger-trees as in DLB07 (i.e. the
Milli-Millennium simulation, with box size of 62.5 Mpc $h^{-1}$).
We use recipes according to our basic formalism presented in
section \ref{sec:formalism}, with details as explained in the
previous sections. We note that occasionally the mass of a
component in a galaxy gets negative when applying the
recipes above. In these cases we set the negative mass to zero as is
done in DLB07.

We compare stellar mass functions from the original DLB07 to the
galaxies in our simplified model in Fig.~\ref{fig:gsmf_DLB07}. For
most masses and redshifts the stellar mass functions are very
similar. One noticeable difference can be seen around
$10^{11}\,\msun$, where our model has a peak at all redshifts. This
peak might be related to the fact that we use the median value of
the cooling efficiency to represent a complex distribution.

\begin{figure}
\centerline{ \hbox{ \epsfig{file=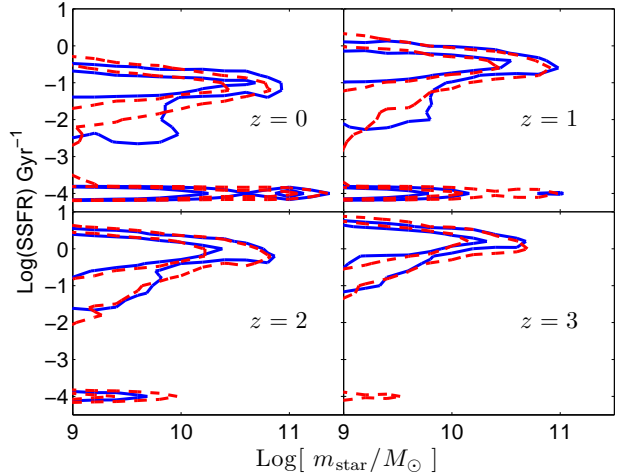,width=9cm} }}
\caption{Specific SF rate (SSFR) versus stellar mass for various redshifts as indicated.
Model 0 galaxies are plotted in solid lines, DLB07 galaxies are plotted in dashed lines.
Contour levels are at 0.01 and 0.1, when the normalization is such that the integral
over all values gives unity. A minimum SSFR of $10^{-4}$ Gyr$^{-1}$ is
used for all galaxies in this plot.}
  \label{fig:ssfr_DLB07}
\end{figure}

\begin{figure}
\centerline{ \hbox{ \epsfig{file=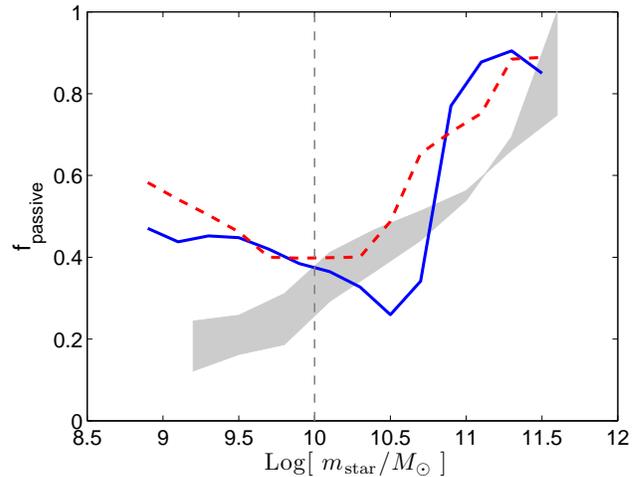,width=9cm} }}
\caption{Passive fraction of galaxies at $\z=0$. Solid line shows Model 0 galaxies,
dashed line shows DLB07. Passive galaxies are defined to
have SSFR lower than the dashed line in Fig.~\ref{fig:ssfr} panel I,
which is also defined in Eq.~\ref{eq:passive_def}.
Observational data is plotted in shaded gray region and is taken from
\citet{Salim07,Schiminovich07}. }
  \label{fig:passive_frac_DLB07}
\end{figure}

In Fig.~\ref{fig:ssfr_DLB07} we compare specific SF rates for
galaxies in our model to those from the DLB07 model. The results are
very similar.
Interestingly, the \emph{scatter} in SF rate for a given stellar mass is
nicely reproduced by our model. This shows that most of the scatter in the properties
of galaxies is not introduced by the scatter in the efficiencies of
the underlying processes, but instead caused by the large variations in
the formation histories of dark-matter haloes. The fraction of passive galaxies
is shown in Fig.~\ref{fig:passive_frac_DLB07} and also follows
DLB07 reasonably well. We note that the results
of SF rates improve for higher redshifts.
Lastly, we show in Fig.~\ref{fig:gas_frac_DLB07} the average fraction of
cold gas in Model 0 versus DLB07. The agreement is again
reasonable, with the exception of the feature visible at a stellar mass
of $10^{11}\,\msun$. This feature is likely related to the
feature in the stellar mass function discussed above.

All the results plotted here show that our model galaxies are very
similar to the galaxies produced by DLB07. Apparently, using
efficiencies which depend on halo mass and time does not seem to
reduce the level of complexity of the model. There are a few
deviations between the models, which is not unexpected given the
extent to which we have simplified DLB07.  Note that
it is likely that our simplified model could
be brought to even closer agreement with DLB07 if
we allowed some freedom in choosing cooling rates, i.e. use
cooling rates which deviate slightly from the median of the
distribution.
Indeed, in section \ref{sec:models} we show
that we can obtain smooth stellar mass functions when using a single
value for the cooling efficiency in each stellar mass and time bin;
the features seen in the stellar mass function and cold gas fraction
at  $10^{11}\,\msun$ are thus not related to the fact that we do not
use a bimodal cooling efficiency like DLB07 do.

\begin{figure}
\centerline{ \hbox{ \epsfig{file=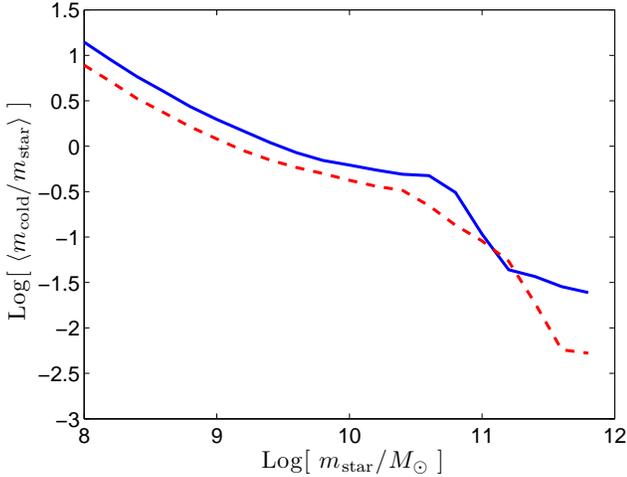,width=9cm} }}
\caption{Average gas fractions, $\mc/\ms$, for Model 0 (solid) and DLB07
(dashed) galaxies. We plot the log of the average over all galaxies at
redshift zero, for a given stellar mass.}
  \label{fig:gas_frac_DLB07}
\end{figure}

\section{The Models}
\label{sec:models}

We use the formalism presented in
section \ref{sec:formalism} and develop specific models that will be able to produce
galaxies which agree well with the observational data. In this section, we present
five models that represent a large range of scenarios.
We discuss the motivation for each model,
and the various ingredients being used. We compare the results of
the models to observation in section \ref{sec:results}.
It should be noted that our tuning is not meant to be perfect, and it
might be that other models similar to ours produce galaxies which fit observations
in a better way.
A summary of all the models developed here can be found in Table \ref{tab:models}.

\subsection{Default ingredients}
\label{sec:default_ingredients}

\begin{table*}
\caption{A summary of the new models discussed in this work. The
models are presented in section \ref{sec:models}, and in
Figs.~\ref{fig:cool_effic} \& \ref{fig:sf_effic}. The results are
summarized in section \ref{sec:results}, Figs.~\ref{fig:mass_funs}
-- \ref{fig:gas_frac}. Components which are considered as free parameters, and used to tune the model
against observations are written on a gray background. }
\begin{center}
\begin{tabular}{p{1cm} || p{2cm} | p{2cm} | p{4cm} | p{3cm} | p{3cm}}
Model & I & II & III & IV & V   \\
 &   zero SN feedback & standard & cold accretion &
only bursts & shut-down by mergers \\
\hline
\hline
$\fc$ & \tune Fig.~\ref{fig:cool_effic} & \tune Fig.~\ref{fig:cool_effic} &
Fig.~\ref{fig:cool_effic}:  $\qquad\qquad\qquad\qquad\qquad\quad$
$\fc \sim \tau_{\rm dyn}^{-1}$ if $\!M_h \lesssim \!(1+\z)\times\!10^{11}\!$ $\hmsun$, else $\fc=0$
 &  same as III & Fig.~\ref{fig:cool_effic}: \tune similar to II, but increased for high mass haloes  \\
\hline
$\fs$ & \tune Fig.~\ref{fig:sf_effic}  &
Eq.~\ref{eq:fsdb07} (see also Fig.~\ref{fig:sf_effic}) &
Eq.~\ref{eq:fsdb07},  in addition: $\;\;\;\;\;\;\;\;\;\;\;\;\;\;\;\;\;$ $\fs=0$ if $M_h>3\times10^{12}$ $\hmsun$ and $\z<0.77$ & 0 &
Eq.~\ref{eq:fsdb07} (see also Fig.~\ref{fig:sf_effic})  \\
\hline
$\ffd$ & 0 & 3.5 & \tune Eq.~\ref{eq:fd_cold}: $\;\;\;\;\;\;\;\;\;\;\;\;\;\;\;\;\;\;\;\;$ $\ffd=10^3 \cdot M_{10}^{-2}$ and not higher than 100& 0 & 3.5 \\
\hline
& & & & \tune & \\
burst \,\,\,efficiency & 0 & $0.56\left(m_1/m_2\right)^{0.7}$ &
$0.56\left(m_1/m_2\right)^{0.7}$ & \tune $0.8\left(m_1/m_2\right)^{0.7}$ for $M_h<10^{12}\hmsun$, otherwise zero &
$0.56\left(m_1/m_2\right)^{0.7}$ \\
\hline
others & -- & -- & $\fe=\ffd$ & \tune Eqs.~\ref{eq:alpha_df} \& \ref{eq:sigma_burst}: $\;\;\;\;\;$ $\sigma_{\rm burst}=5(t/13.6)^2$ $\alpha_{\rm df}=(t/13.6)^2$ & \tune shut-down of cooling \& SF after mergers of mass ratio $>$0.2 \\
\hline
section & \ref{sec:model1} & \ref{sec:model2} & \ref{sec:model3} &
\ref{sec:model4}& \ref{sec:model5} \\
\hline
line-type & thin blue & thin dashed red & dotted-dashed green & dotted brown
& thick dashed pink \\
\end{tabular}
\end{center}
\label{tab:models}
\end{table*}

We list below the default ingredients of the models. These are common to
most of the models discussed in this work. Deviations from these basic
ingredients will be
addressed specifically for individual models.
\begin{itemize}
\item Hot gas that is being stripped from satellite galaxies is lost to the
intergalactic medium and is not added to the hot gas reservoir of the
central galaxy. Although other SAMs usually add the stripped gas to the central
galaxy, it might be that tidal interactions will eject some of this gas out of the halo.
Also, this effect could be counterbalanced by changing cooling rates for high-mass haloes.
We use $\alpha_h=0.25$
Gyr$^{-1}$ which mimics a stripping with an exponential time-scale of 4 Gyr for all satellite galaxies.
\item The dynamical friction time $t_{\rm df}$ from Eq.~\ref{eq:t_df} is
multiplied by a factor of $\alpha_{\rm df}=3$.
\item  The efficiency of merger-induced SF bursts is given by $\alpha_b=0.56,\; \alpha_c=0.7$
as suggested by DLB07. The burst duration, $\sigma_{\rm burst}$, is assumed to be very short, at the level
of 10 Myr.
\item The ejection efficiency, $\fe$, is set to zero.
\item We use a recycling factor of $R=0.5$. This value
is obtained using the \citet{Bruzual03} stellar population synthesis model,
adopting a \citet{Chabrier03} IMF.
\item All gas is accreted to the hot gas phase, so $\fca=0$, $\fha=0.17$.
\end{itemize}

\begin{figure}
\centerline{ \hbox{ \epsfig{file=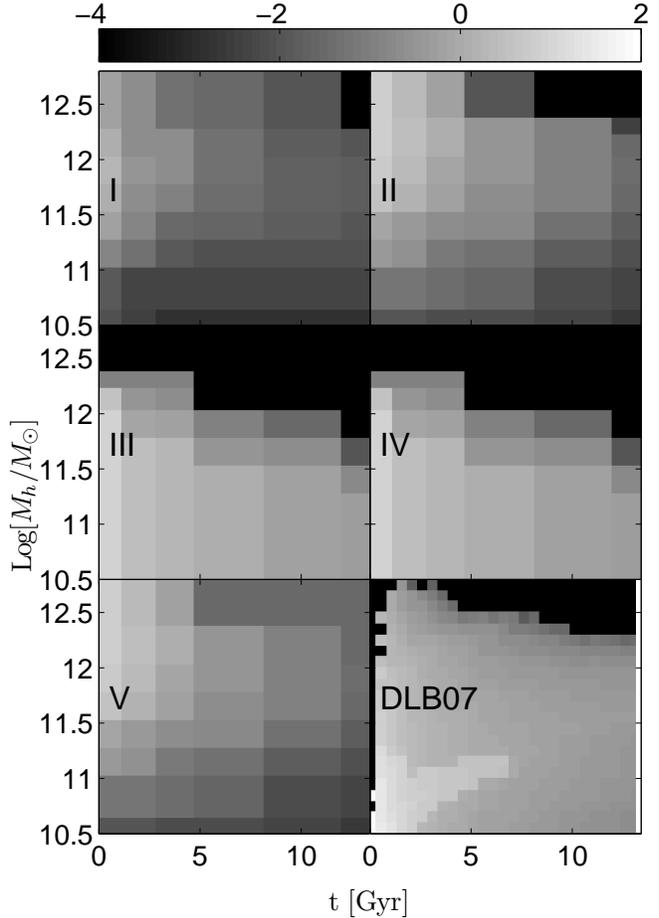,width=9cm} }}
\caption{Cooling efficiencies used in our various models.
The gray scale shows Log  values  of  $\fc$ in units of Log[Gyr$^{-1}$].
X-axis represents the time in Gyr since the big-bang, y-axis is the host halo mass.
Each panel is devoted to a different model
as indicated (see summary of all the models in table \ref{tab:models}).
For haloes more massive than shown here we use the same
efficiencies as for the most massive bin.
Cooling efficiencies in the bottom right panel were obtained by
running the SAM of \citet{DeLucia07} and are discussed in section
\ref{sec:test}. We use a bilinear interpolation scheme to obtain logarithm values
for intermediate masses and times. Specific values for the efficiencies plotted here can
be found in Appendix A.}
\label{fig:cool_effic}
\end{figure}

\subsection{Model I -- zero SN feedback}
\label{sec:model1}

In section \ref{sec:cooling} we argue that the efficiency of cooling
might be different from the values predicted by \citet{White91}.
Here we would like to test a scenario in which this efficiency has
its minimum possible value. It will be interesting to have a
constraint on the minimum cooling rates from observations. We can
then examine if these values fall within the range of uncertainties
discussed in section \ref{sec:cooling}.

A main degeneracy in modeling the evolution of galaxies is
that cooling can be balanced by feedback processes which heat up the
cold gas. Thus, a minimum cooling scenario must be achieved when the
feedback efficiency is set to zero, while observations are
still reproduced. In this case,
the total mass in cold gas and stars depends only on the cooling efficiency.
Consequently, if the total mass of stars and cold gas is known for a statistical
sample of galaxies, we can define the minimum cooling rates in a relatively unique
way. Unfortunately, we only have observational estimates on cold
gas fractions at very low redshifts. Nevertheless, we find that
those are already a significant constraint, as they limit
the sum of cooling efficiencies over all times. We set the efficiency of
merger-induced bursts to zero, in order to further increase the
degree of uniqueness in this problem.

We tune the cooling and SF efficiencies iteratively. Once
we have assumed certain cooling efficiencies, we tune the SF efficiencies to get the proper stellar mass
functions. If the resulting
amount of cold gas differs from observations we then change the cooling efficiencies again,
and re-tune SF accordingly. Only a few iterations are needed in order to converge to the solution given below.

The cooling and SF efficiencies of this model are shown in Figs.~\ref{fig:cool_effic} \&
\ref{fig:sf_effic}. As expected, cooling efficiencies in this model
need to be very low, and are roughly 1-3 orders of
magnitude smaller than the standard values adopted by the usual
SAMs (values from the SAM of DLB07 are shown in the same
figure). Such a low efficiency might be explained by substantial
preheating.
For high redshift and massive haloes ($\z>1$, $M_h\gtrsim
10^{12}\hmsun$) the cooling efficiencies used in Model I are only roughly
an order of magnitude lower than the standard ones, which might be a
plausible difference relative to the standard cooling rates
even in the absence of preheating. An additional
characteristic of cooling efficiencies is that they are peaked at a
halo mass of $\sim6\times10^{11}\hmsun$ at all redshifts.

\begin{figure}
\centerline{ \hbox{ \epsfig{file=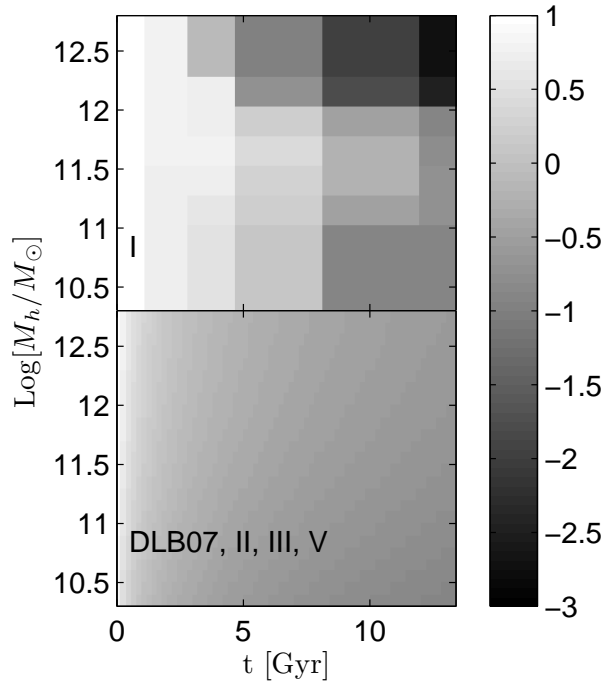,width=9cm} }}
\caption{SF efficiencies values ($\fs$) used in our various models
in units of Log[ Gyr$^{-1}$ ]. This figure is similar
to Fig.~\ref{fig:cool_effic}. The functional shape of  DLB07
is given in Eq.~\ref{eq:fsdb07}. Specific values used by Model I are given
in Appendix A.}
\label{fig:sf_effic}
\end{figure}

Interestingly, we can see in Fig.~\ref{fig:sf_effic} that
our zero feedback model requires a star formation efficiency much higher
than  in  DLB07  at high redshifts.  This  is necessary in order to  obtain
reasonable  cold  gas  mass  fractions  at $\z=0$: If  star  formation
efficiencies were brought down, this would
have to be  counterbalanced by higher cooling rates  at high redshift,
which would in turn lead to too high gas fractions at $\z=0$.


\subsection{Model II -- standard}
\label{sec:model2}

In Model II we adopt feedback and SF efficiencies which are similar to
that of DLB07 (see section \ref{sec:test}). However, we
differ from DLB07 in three respects: (i) we do not include an ejected
phase, (ii) we do not include threshold for SF, and (iii)
the hot gas of satellite is stripped exponentially
with a relatively long time-scale of 4 Gyr. We then look for
cooling efficiencies which produce galaxies that
fit the observational data given the above assumptions.

As in Model I, cooling rates within Model II have to be low in
order  to reproduce  observations. These are shown in Fig.~\ref{fig:cool_effic}.
Compared to DLB07  cooling rates are lower by up to two  orders of magnitudes
(instead of the three order of magnitudes in  Model I). This
means that the mechanism of ejection  actually has  a stronger
impact than  the  pure `reheating' by supernova feedback; reheating by
supernova seems to allow to increase cooling rates  by one  order of
magnitude;  ejection by two  orders of magnitude.

\begin{figure*}
\centerline{\psfig{file=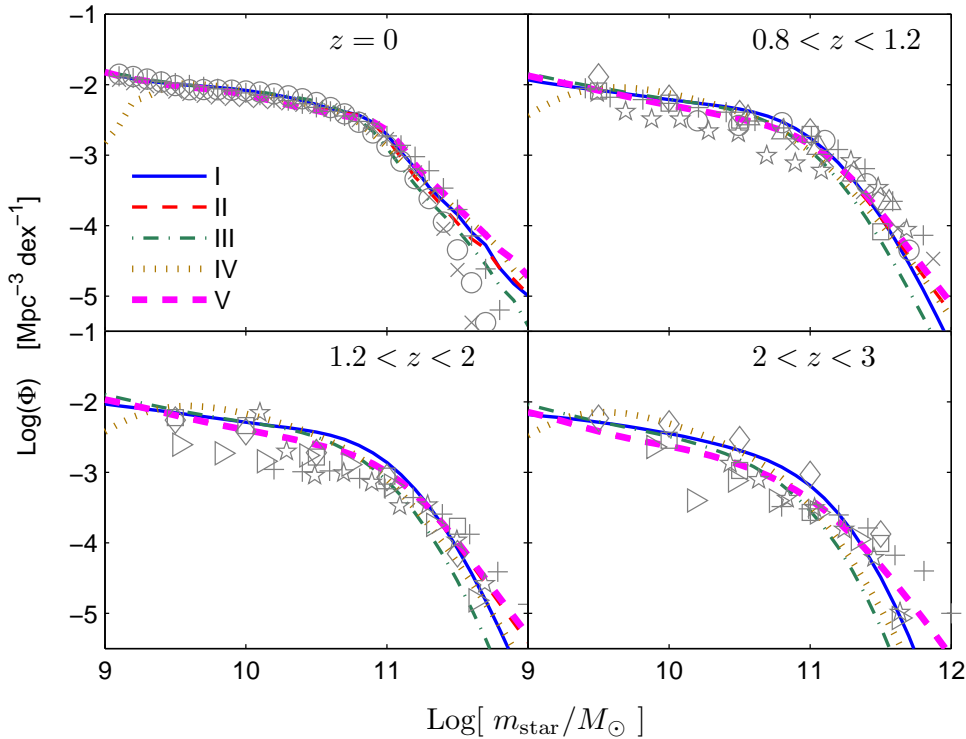,width=150mm,bbllx=30mm,bblly=80mm,bburx=188mm,bbury=200mm,clip=}}
\caption{Stellar mass functions of galaxies at various redshifts, for our different models as indicated in
the top left panel and as summarized in table \ref{tab:models}. The
observational data is plotted with gray symbols. At all redshifts
higher than zero we convolve the model stellar masses with a
Gaussian error distribution, with standard deviation of 0.25 dex. At
$\z=0$ we use observations by \citet[][circles]{Li09}, \citet[][crosses]{Baldry08},
\citet[][pluses]{Panter07}. At high-$\z$ we use the following
observations: \citet[][$\z=0.75-1$, circles]{Bundy06},
\citet[][$\z=0.8-1$, crosses]{Borch06}, \citet[][$\z=0.8-1$,
$\z=1.6-2$, $\z=2.5-3$, plus signs]{PerzGonzalez08},
\citet[][$\z=0.8-1$, $\z=1.6-2$, $\z=2-3$, stars]{Fontana06},
\citet[][$\z=0.8-1$, upward-pointing triangles]{Drory04},
\citet[][$\z=0.75-1.25$, $\z=1.75-2.25$, $z=2.25-3$, diamonds and
squares]{Drory05}, \citet[][$\z=1.3-2$, $\z=2-3$, right-pointing
triangles]{Marchesini09}. Model stellar mass functions are plotted
at $\z=0$, 1, 1.5, 2.5 according to the label on each panel. We
treat the specific IMF chosen in each measurement as part of the
observational `uncertainty' and do not convert them into
the same IMF. } \label{fig:mass_funs}
\end{figure*}


\subsection{Model III -- cold accretion}
\label{sec:model3}

In Model III we include a prescription  for cold accretion
which follows the ideas described by \citet{Birnboim03}, \citet{Cattaneo06},
\citet{Keres09} and \citet{Khochfar09}.
In order to model the infall of cold gas into the disk, we assume that $\mh$ represents
the mass within \emph{cold filaments} flowing with virial velocity into the disk.
In this case, the rate of transition from $\mh$ to $\mc$ is fixed by the infall time,
which we approximate with
the dynamical time of the halo. Consequently, the `cooling' efficiency is now:
\begin{equation}
f_{c,{\rm max}} = 1/\tau_{\rm dyn} \approx \frac{1}{0.15t}\,.
\end{equation}
This sets the rate at which gas that joins the halo reaches the disk and becomes
available for star formation.
 This is also the \emph{maximum} cooling efficiency that can be used. From
Fig.~\ref{fig:cool_effic} it can be seen that the cooling efficiencies used by DLB07
in the ``cold accretion regime'' are slightly
lower (usually by 0.2 dex, 60\%) than the maximum efficiency $f_{c,{\rm max}}$
defined here.

We roughly adopt the results of \citet{Keres09} which indicate that cold accretion is dominant
at halo masses below few times $10^{11} \hmsun$, and use maximum cooling efficiencies below
a mass of $\sim(1+\z)\times10^{11}\hmsun$ as can be seen from Fig.~\ref{fig:cool_effic}.
We use the same SF efficiency  as in  DLB07  (see section \ref{sec:test} and
Fig.~\ref{fig:sf_effic}), except that  we set the SF
efficiency to zero for haloes more massive than $3\times10^{12}\,\hmsun$ at redshifts
smaller than 0.77, in order  to produce the proper number of passive and massive galaxies.
This may mimic an AGN feedback in massive haloes which heats up
all the cold gas in the disk, and does not allow any additional accretion of cold gas
coming from satellite galaxies.

Both cooling and SF efficiencies are thus set by our preliminary assumptions.
We are now looking for a SN feedback efficiency that will make this model successful in reproducing
the population of galaxies. In order not to confuse the heated gas due to feedback with
$\mh$ which is actually the filament mass here, we use the ejection efficiency $\fe$ and
remove all the heated gas from the halo.
We find that a simple power-law gives reasonable results:
\begin{equation}
\fe=\ffd= 10^3 \cdot M_{10}^{-2} \,,
\label{eq:fd_cold}
\end{equation}
where $M_{10}$ is the subhalo mass $M_h$ in units of ${10^{10}
\hmsun}$.  We limit this efficiency to a maximum value of a 100 Gyr$^{-1}$
in order for the code to converge within 20 time-steps, as discussed
at the end of section \ref{sec:usage}.


\subsection{Model IV -- only bursts}
\label{sec:model4}

The star formation  histories  of  galaxies  are a  combination  of  short
episodes of bursts and long term quiescent  evolution. These two
processes are triggered by different mechanisms originating from the
dark-matter  merger history.  Bursts  are assumed  to follow  mergers,
while  quiescent evolution  is due  to  smooth accretion  of gas  into
galaxies. The relative contribution of these two mechanisms to the
total stellar mass formed is observationally only partially constrained
(for example, the total mass of stars within thin disks at low redshift
might give an estimate for the total quiescent SF history).
Here, we investigate if a model where stars are made in bursts can
be  distinguished  from  the  other  more standard  models.   We  will
therefore construct an extreme model in which all
stars are formed in SF bursts, hoping that it will
reveal some  of the basic differences  between
the  merger and the quiescent mode of SF. These  differences may
then be used  in future work to distinguish  between more reasonable models,
where bursts contribute much less to the overall SF
history  of the  universe. \citet{Noeske07}, for example, use the
scatter in the SF rates of galaxies at $\z=1$ to infer that starburst
in gas-rich major merger are not a dominant mode of star formation
at these redshifts.

We set the feedback efficiency and quiescent mode of SF to zero.
Cooling efficiencies are adopted from the cold accretion model. The only freedom
in this model is in fixing the dynamical-friction time, the burst duration, and burst efficiency.
We find that a good match to observational data is obtained if we use
\begin{equation}
\alpha_{\rm df}= \left[\frac{t}{13.6}\right]^2 \,,
\label{eq:alpha_df}
\end{equation}
for the dynamical-friction time (Eq.~\ref{eq:t_df}). We assume the burst dependence
on time has a Gaussian shape, with a standard deviation of
\begin{equation}
\sigma_{\rm burst} = 5 \left[ \frac{t}{13.6} \right]^2 \;\; {\rm Gyr}\,.
\label{eq:sigma_burst}
\end{equation}
The peak of the SF efficiency is defined to happen at $2\sigma_{\rm burst}$
after the galaxies coalesce. The reader is referred to section
\ref{sec:df_uncertainties} for a discussion about uncertainties in the above
recipes.
Lastly, we modified the burst efficiency and use $\alpha_b=0.8$, $\alpha_c=0.7$
only for haloes below a mass of $10^{12}\hmsun$. More massive haloes are not
allowed to have any merger-induced bursts.


\subsection{Model V -- shutdown by mergers}
\label{sec:model5}

Different scenarios for the build-up of red galaxies (the so-called
`red sequence') are still under debate. Four main
channels for galaxies to become red have been suggested:
(i) major mergers can trigger
bursts of SF, exhaust the gas reservoir, and trigger AGN activity
\citep{Toomre72,Mihos94,Hopkins08}; (ii) above some halo
mass cooling rates may be decreased due to AGN feedback or virial shocks
\citep{Birnboim03, Dekel06, Cattaneo06, Keres05, Bower06, Croton06,
Somerville08}; (iii) morphological quenching \citep{Martig09}
(iv) environmental  effects acting on satellite galaxies.

Here we present a model in which major mergers completely shut off
both cooling and star formation for all time thereafter, which could be explained
by a combination of scenarios (i) and (iii) above.
This is in contrast to the previous models in which SF and cooling is quenched
above a given halo mass (see e.g. Fig.~\ref{fig:cool_effic})

This model is similar to the model by \citet{Hopkins08}
who have  suggested that major mergers  of gas-rich
systems  may lead  to permanent quenching  of star  formation.
These authors compared
their results to the standard semi-analytical models \citep[e.g
DLB07;][]{Cattaneo06,Somerville08} which basically assume that
star formation  is quenched above some  fixed halo mass. They claim that their
model leads to an improved agreement with  observations for
(i)  the passive fraction  as  a function  of stellar  mass and halo
mass (ii)  the number  of massive  and passive galaxies at high
redshift.   Here we try out a
similar model. We start from Model II (i.e. similar to DLB07),
but assume that any galaxy undergoing a major merger (with a mass
ratio below 5:1) has its cooling and star formation efficiency set
to zero permanently. Not even merger-induced bursts are allowed
anymore. This makes it possible to increase cooling rates
substantially for massive haloes, as visible in Fig.~\ref{fig:cool_effic}. Regarding
all other recipes, this model is equivalent to Model II.


\section{Results}
\label{sec:results}

In this section we discuss the results obtained from the five models
presented in the previous section using the full Millennium simulation. When tuning the models we examine the
same set of figures as seen here for each model run. In the first part of the
tuning we try to match
the stellar mass functions and the universal SF density plots.
As a second priority we consider the distribution of SF rates, the cold
gas fractions, and the passive fraction of galaxies at $\z=0$.
It might be that our models depend strongly on the priorities we give
different observational constraints. It might also be that within
the assumptions of each model there are other possible models
providing equal or better results. The main goal in tuning these models
is to point out the interplay between different recipes, and to establish
a set of models for future studies (e.g., merger rates, bimodality in SF).
We do not claim to provide an excellent match to the observational results.
However, the majority of our models give similarly good agreement with
observations as other current SAMs.

\subsection{Stellar masses}
\label{sec:results_stellar_mass}

\begin{figure}
\centerline{ \hbox{ \epsfig{file=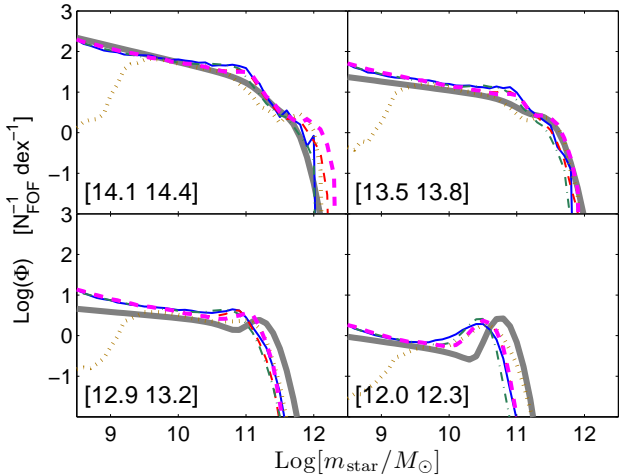,width=9cm} }}
\caption{The conditional stellar mass functions, for a given \fof
halo mass at $\z=0$. Model results are plotted
according to line-styles which are given in table \ref{tab:models}. The data
from \citet{Yang09} is plotted in thick gray lines. In each panel we quote
the relevant range in halo mass in units of logarithm of $\hmsun$ according to
\citet{Yang09}.  We transform these ranges to the halo mass used by the Millennium simulation
according to equation 2 from \citet{Weinmann06a}.}
  \label{fig:halo_cmf}
\end{figure}

The stellar mass function is the most basic physical constraint on the population
of galaxies. Observationally, measuring stellar masses for galaxies is not a
trivial task. It involves various steps with non-negligible uncertainties.
The transformation of photometrical data to stellar masses, the assumed IMF,
dust corrections and cosmic variance may all contribute to significant deviations
between different studies \citep[see for example the discussion in][]{Baldry08}.

In  Fig.~\ref{fig:mass_funs} we plot stellar mass functions
for all our models against observations in various redshift bins.
It can  clearly be seen  that all
our models fit the stellar mass  functions relatively accurately, and, most importantly all models
have almost the \emph{same} mass functions. This indicates that although the stellar mass function over
time constrains the parameters of the models, it can  be
reproduced rather easily  with  very  different basic scenarios.
There are few deviations between our model results and observational data,
mainly at high $\z$ and for low mass galaxies. In addition, Models IV and V slightly
overpredict the high  mass end of the stellar mass function at $\z=0$. The overall consistency
 is surprising given  the large differences between the models.

The assumption about the shape of the stellar initial mass function (IMF)
is an important ingredient in the observational analysis. However,
our model does not include an assumption about the IMF except for the relatively
small dependence on the recycling factor, $R$.
Consequently, we do not attempt to correct the data to the same IMF and we
see it as part of the observational uncertainty.
It might be that only parts of the data sets are fitted properly by our model, indicating
that some specific IMF solutions are preferred. We do not try to explore this issue further
here. Changes in the IMF between different observational results might appear as
inconsistencies within the data (compare the high-mass end at $\z=0.8-1.2$ versus $\z=0$).
These issues are less important here, as we are mainly concerned with comparing our
models to different observational predictions.

The stellar mass function obtained from Model IV indicates a severe resolution
problem below $\sim5\times10^9\,\msun$. Only this model
shows such a problem which is connected to the fact
that contributions from mergers are much
more sensitive to the merger-tree resolution than other processes.
This indicates that the minimum galaxy mass which is properly resolved in our model
strongly depends on the actual parameters being used.

More detailed information on the stellar masses of galaxies is obtained
by splitting the mass function into galaxies which live inside different
halo masses. In Fig.~\ref{fig:halo_cmf}, we  compare the conditional  stellar
mass function for a given halo mass  obtained  with  our models  to  the  one  given by
\citet{Yang09}.   Agreement with observations  seems to be
reasonable for all our models,  except in the lowest halo mass  bin,
where our models predict a  lower stellar  mass to halo  mass
relation for  the central galaxies. These deviations might be due to the limited accuracy to which
the halo mass can be measured, or due to the different
cosmological parameters used by \citet{Yang09} and by the Millennium simulation.
We also note that the somewhat too high number of small
mass satellites at $\z=0$ may be correlated to the slight
overprediction of the stellar mass functions
at high-$\z$. This is because satellite galaxies today were central galaxies at high-$\z$
and thus were dominating the stellar mass functions at high-$\z$.

\subsection{Star formation rates}
\label{sec:results_sfrs}

\begin{figure}
\centerline{ \hbox{ \epsfig{file=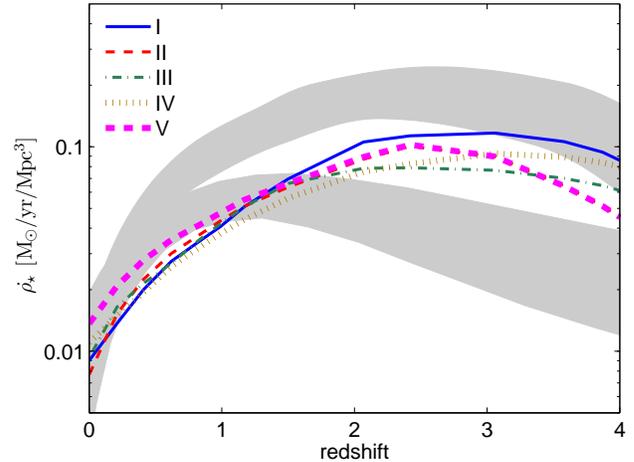,width=9cm} }}
\caption{The universal SF rate density. Model line styles are described in
table \ref{tab:models}. Gray shaded regions show observational data:
the upper region is taken from the compilation by \citet[][3$\sigma$ confidence level]{Hopkins06};
the lower region is taken from \citet[][1$\sigma$ confidence level]{Wilkins08}
and corresponds to the universal SF rate density derived from the evolution of
the stellar mass functions with $\z$.}
  \label{fig:madau}
\end{figure}

\begin{figure}
\centerline{ \hbox{ \epsfig{file=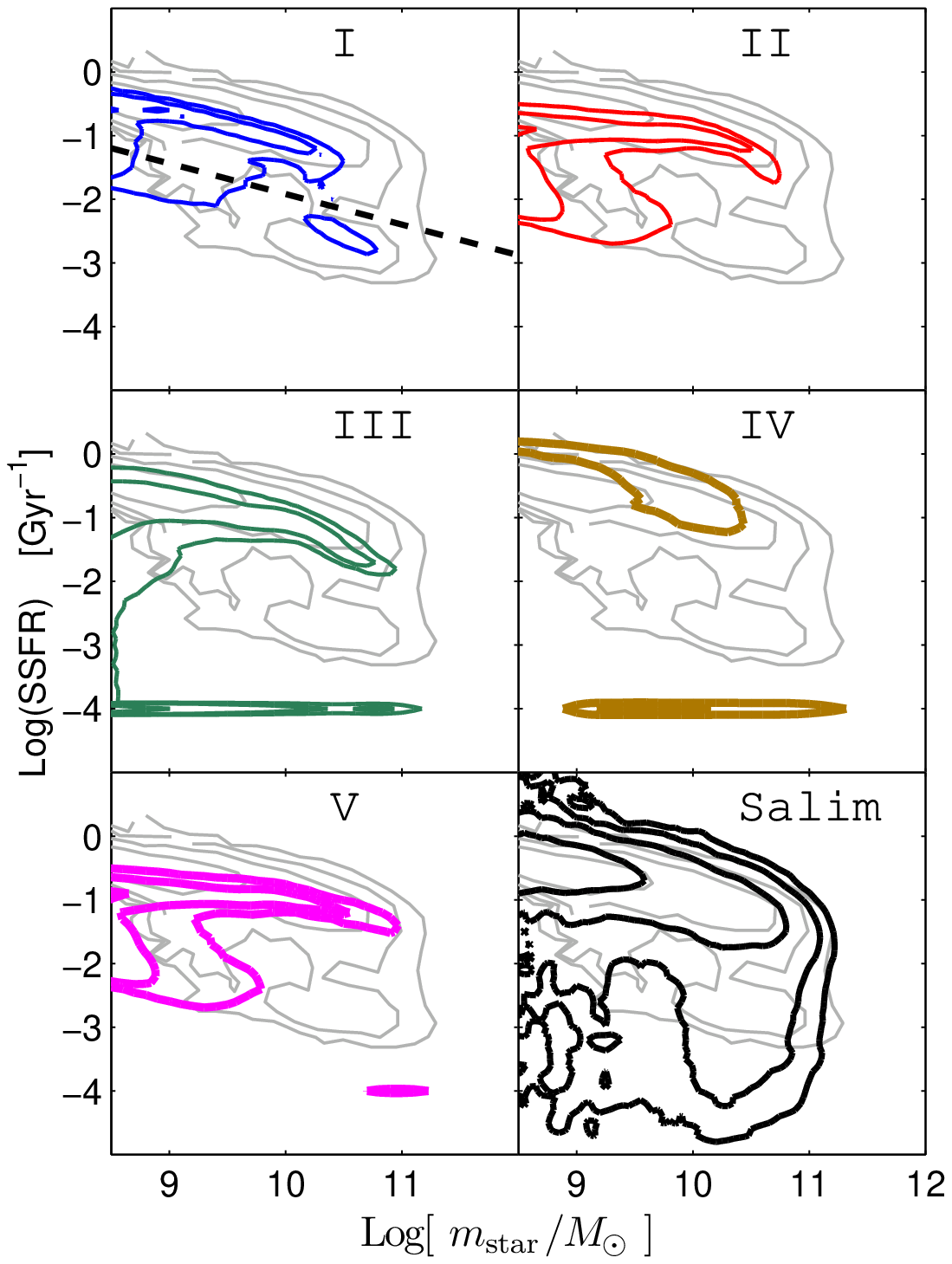,width=9cm} }}
\caption{The two-dimensional distribution of specific SF rate (SSFR) versus stellar
mass at $\z=0$. Observational data were graphically extracted
from \citet{Schiminovich07} and are plotted in gray lines. Model results are plotted
in solid thick lines. The data from \citet{Salim07} is plotted over
\citet{Schiminovich07} in the bottom right panel,
 in order to demonstrate the importance of observational
uncertainties (we slightly smooth the data from \citet{Salim07} in order to make
the contours readable, this does not change the contour average location).
In all panels contours show the regions which encompass
 38\%, 68\%, 87\% and 95\% of galaxies.
We use a minimum SSFR value of $10^{-4}$ for all galaxies in this plot, as observations
are not sensitive to lower SSFR.
We define galaxies as being `passive' if their SSFR falls
below the dashed line shown in panel $I$. This line is a rough estimate
drawn by eye, and is given in Eq.~\ref{eq:passive_def}.}
  \label{fig:ssfr}
\end{figure}

In  Fig.~\ref{fig:madau},  we  show  the  universal SF rate  density  as
a function of  redshift. Again,  it is interesting  that all  our
models seem  to  be   relatively  similar, and they  all
lie within the observational uncertainties. We show two kinds
of observational results here. The first comes  from directly
measuring  the SF rate \citep{Hopkins06}, and the second
from differentiating the stellar mass
density of the universe with respect to time \citep{Wilkins08}.
We have not managed to
simultaneously  reproduce  the stellar  mass functions over  time
and the  high SF rate density found in direct measurements.
This inconsistency was already noted by e.g. \citet{Hopkins06,Wilkins08}.

\begin{figure*}
\centerline{\psfig{file=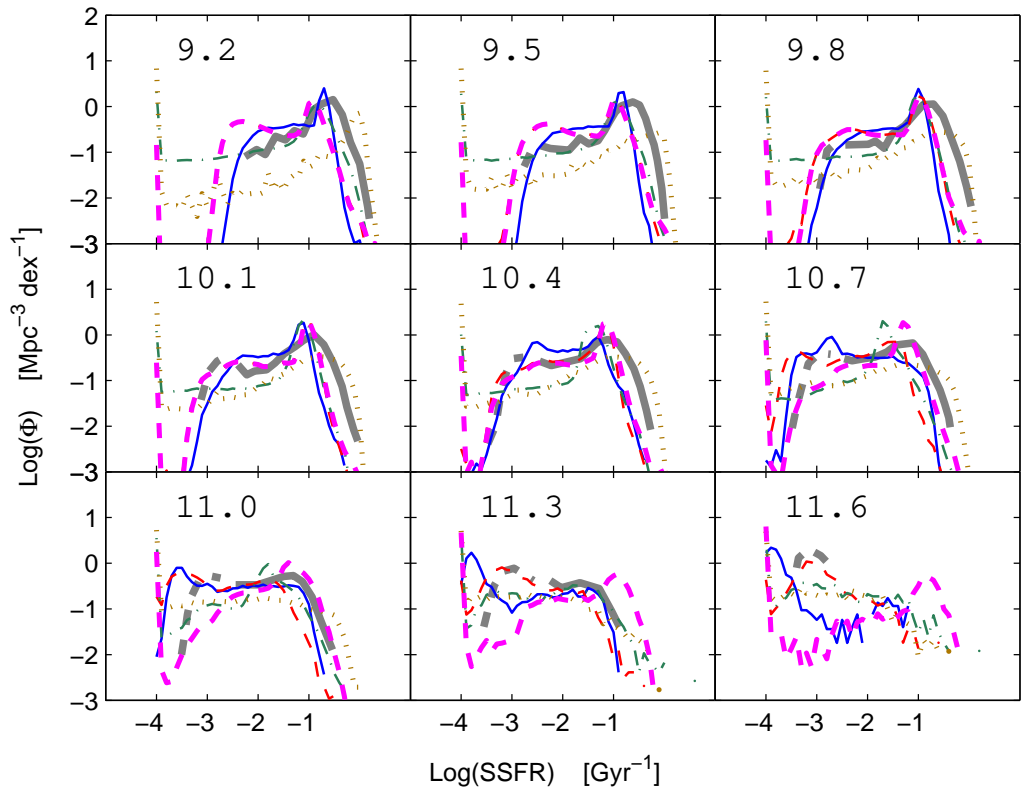,width=150mm,bbllx=30mm,bblly=80mm,bburx=188mm,
bbury=200mm,clip=}} \caption{The probability  distribution of
specific SF rates (SSFR) for different bins in stellar mass at $\z=0$.
Average stellar mass in Log $\msun$ is indicated at the top of each
panel. Observational results from \citet{Schiminovich07} are shown
in thick gray lines. For SSFR values below $10^{-2.5}$ Gyr$^{-1}$ we
plot the observation in dashed lines in order to emphasize the fact
that these estimates are highly uncertain. Results for the model
galaxies are plotted according to table \ref{tab:models}. We use a
minimum SSFR value of $10^{-4}$ for all the model galaxies.
In order to emphasize the shape of the distributions we normalized all
lines to represent the same number of galaxies within a given mass bin.}
\label{fig:ssfr_cuts}
\end{figure*}

In Fig.~\ref{fig:ssfr}  we show the distribution of  the logarithm
of the specific SF rate (defined as the SF rate divided by the stellar
mass; hereafter SSFR) as a function of stellar mass at $\z=0$. Here we start
to see larger deviations  between our different models,  indicating
that this relation is much more constraining  than the stellar mass
functions or the  SF rate density of  the  universe.  Unfortunately,
observations  are relatively uncertain, especially for  passive
galaxies \citep[see the discussion in][]{Salim07}.
Note that both observational data sets use volume-corrected values.

Interestingly, all our models  manage to reproduce the tilt of the
star-forming sequence,  which is not  reproduced by  other current
SAMs we are  aware of \citep{Somerville08,Fontanot09}.
This is somewhat surprising, as our models were not specifically tuned
to reproduce this tilt. We suspect that the absence of the tilt in most standard
SAMs is related to the inclusion of the ejection phase, as our Model 0
from section \ref{sec:test} which mimics
DLB07 does not reproduce the tilt either. We have checked
that removing the process of ejection in Model 0 results
in the appearance of a tilt in this model, although the stellar mass function is no longer similar
to observations. In order to correct the stellar mass function we need to modify the cooling
efficiencies, as is done in Model II. We plan to further explore the SSFR versus
stellar mass diagram, and the origin of the tilt in the star-forming
sequence, in a future study.

There are a few differences seen in Fig.~\ref{fig:ssfr} between the observed and model
distributions which are somewhat artificial and do not represent important discrepancies.
Uncertainties in measuring stellar masses and SSFR can be modeled by multiplying our model
results by a random error distribution. This will make the 2-dimensional distribution
wider, and will probably better fit the observations results. In addition,
we do not allow SSFR to go below 10$^{-4}$ Gyr$^{-1}$ in the diagram. This results in a strong
artificial peak at this SSFR value. Such a peak is not seen in the observational data,
where passive galaxies may have SSFR values smeared up to few times 10$^{-3}$ Gyr$^{-1}$.

In Fig.~\ref{fig:ssfr_cuts} we show  the distribution of SSFR
for all our  models, in 9 different stellar
mass bins, at $\z=0$.
This figure shows that although the general tilt seen in Fig.~\ref{fig:ssfr} is reproduced,
for low stellar masses below $10^{10}\,\msun$ our model star-forming galaxies have lower SSFR
than the observed ones with a difference of $\sim0.3$ dex.
The SSFR values for passive galaxies (lower than $\sim10^{-3}$ Gyr$^{-1}$) are
highly unconstrained by the observations. However, it is interesting that each model shows a very unique
behaviour in this regime. Once observations will be able to provide more accurate estimates of SF
rates for passive galaxies, this information may be
useful in constraining galaxy formation models.
In Appendix B we show a similar figure, but comparing to the results of \citet{Salim07}.

In  Fig.~\ref{fig:passive_frac}  we compare  the fraction  of
passive galaxies as a function of stellar
mass in our five different  models. Passive galaxies
are defined according to the line shown in panel I of Fig.~\ref{fig:ssfr}.
This line was drawn to roughly divide the population of passive versus active galaxies,
and it corresponds to
\begin{equation}
 \log({\rm SSFR_{passive}}) = -0.48 \log \ms + 2.88 \,,
\label{eq:passive_def}
\end{equation}
where $\ms$ is in units of $\msun$.  While  this  observational
quantity  is  rather uncertain, which  is reflected in  the already
quite  large difference between passive fractions derived by
\citet{Salim07}  and \citet{Schiminovich07}
 (which are both  based on the UV-data of the same
set  of galaxies, using different methodologies), it
could be  powerful to  constrain SAMs. In all  our models, except Model IV,
the increase of the  passive fraction as a
function of  stellar mass is  too steep. This  is probably due  to the
fact that  all galaxies in  high mass haloes  are quenched in  our
models in the same way.
At low stellar masses, on the other hand, all our models
overproduce the fraction of passive galaxies significantly.
This might be a result of the difficulty of detecting low mass red
galaxies in the observational data, or the specific treatment of satellite
galaxies in our models. We note that a similar trend is found in the DLB07
model, shown in Fig.~\ref{fig:passive_frac_DLB07}.

\begin{figure}
\centerline{ \hbox{ \epsfig{file=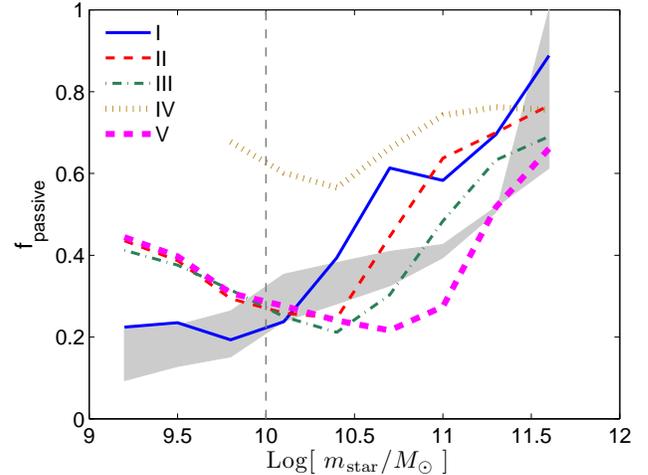,width=9cm} }}
\caption{The fraction of galaxies which are passive out of all the
galaxies within a stellar mass bin at $\z=0$. Passive galaxies are defined to
have SSFR lower than the dashed line in Fig.~\ref{fig:ssfr}, panel I (see also Eq.~\ref{eq:passive_def}).
Observational data is plotted in shaded gray region and is taken from
\citet{Salim07,Schiminovich07}. Results from our various models are plotted
according to the usual line-types.}
  \label{fig:passive_frac}
\end{figure}

\subsection{Cold gas masses}
\label{sec:results_coldgas}

In Fig.~\ref{fig:gas_mf}, we show  the cold gas mass function at
$\z=0$ for  our  different  models. This
observation  provides important additional information about
galaxy formation which can potentially  be used  to discriminate  between
models. Unfortunately, observational constraints are still uncertain
as they only measure the \HI\ gas content. In order to obtain the
mass function of the total cold gas,  the mass of molecular gas
needs to be estimated. This is done here using the model results from
\citet{Obreschkow09a}.
Clearly, both  the zero feedback, and  the ``shutdown  after mergers'' models
tend to  overproduce the  number of objects containing  large
amounts of  cold gas. In both models, a mechanism like for example
AGN feedback that heats up cold
gas above some halo mass, or after a major merger, could solve this
problem; we have not included such a mechanism for simplicity.

Average gas fractions as a function of stellar mass are shown
in Fig.~\ref{fig:gas_frac} along with the observational results from
\citet{Catinella10}. Gas fractions can help us  to break the  degeneracies
between cooling and star formation efficiencies.
Clearly, the gas fractions are much too high in the ``shutdown
after merger'' model; however, observations are still too uncertain
to distinguish between our other scenarios. Future, more  detailed observations of the
cold gas fraction including both \HI\ and molecular gas   will
therefore be an  extremely   important   test  for SAMs.

\begin{figure}
\centerline{ \hbox{ \epsfig{file=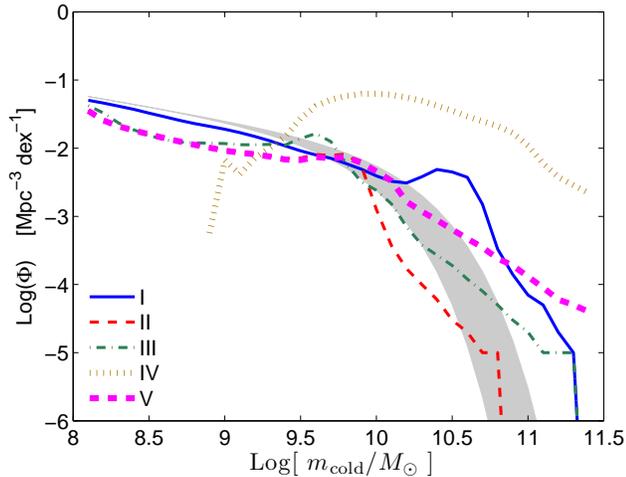,width=9cm} }}
\caption{Cold gas mass function at $\z=0$. Observational data for the \HI\ atomic gas
are taken from \citet{Zwaan05} and define the lower part of the gray shaded region.
In order to estimate the total amount of cold gas (atomic and molecular)
we use the model and predictions
of \citet{Obreschkow09a} plotted as the upper part of the shaded region.
Model results are plotted according to table \ref{tab:models}. }
  \label{fig:gas_mf}
\end{figure}

\begin{figure}
\centerline{ \hbox{ \epsfig{file=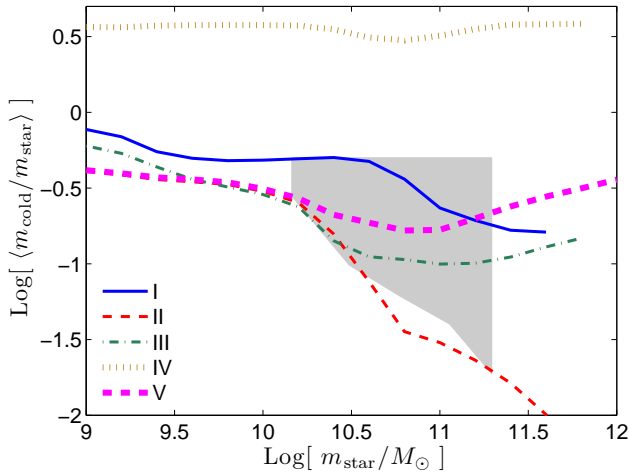,width=9cm} }}
\caption{Average cold gas fraction, $\mc/\ms$ as a
function of stellar mass at $\z=0$.
In order to estimate the observational constraints we use preliminary results
from the GALEX Arecibo SDSS survey \citep[GASS,][]{Catinella10}.
These are estimates for the
average \HI\ content for a complete sample of galaxies, and thus define the lower
limit for the cold gas in our model (atomic and molecular gas). Upper limits are set by
the rough estimates of \citet{Bothwell09} to the \HI/H$_2$ mass ratio and are highly uncertain.}
  \label{fig:gas_frac}
\end{figure}


\section{Summary and Discussion}
\label{sec:discuss}

In this paper we develop a new formalism for the formation and
evolution of galaxies within a hierarchical universe. Our approach
is similar to semi-analytical models (SAMs), although we try to be
more simple and schematic in order to encompass a large range of
possible models.

We claim that a significant step towards more simple, general, and
transparent models is achieved by parameterizing the basic processes of
galaxy formation as functions of host halo mass and redshift alone.
This language enables an easy and compact summary of the model, and
results in a well defined set of differential equations. Once each
recipe is just a function of halo mass and redshift it is also
simple to check the dependencies between recipes, and to determine
 those that will result in a
population of galaxies which matches the observational constraints.

We compare our approach to the SAM of \citet{DeLucia07} and show
that also within this specific model, the recipes are effectively a
function of halo mass and redshift. The only recipe which
deviates from this approximation to some degree
is the radiative gas cooling.
However, for the range where the deviations are important
(i.e. small mass haloes), it is still not
clear how accurate the cooling approach adopted by this SAM is.
Using the median values per halo mass and redshift for each recipe
calculated from the original SAM, our model can produce a very
similar population of galaxies. This demonstrates that forcing the
physical recipes to depend only on halo mass and redshift does not
affect the results of the SAM substantially, at least for the
properties examined here.

The general language we develop here might be useful for comparing
the various SAMs currently in use and for highlighting differences
between models. It can also serve as a tool for comparing
hydrodynamical simulations against SAMs. For example, it might be
possible to obtain average feedback efficiencies as a function of
halo mass and redshift using detailed simulations,
and then to insert this information into the SAMs. This simple
language might help to improve the communication between the different
methodologies.

The approach adopted here is very flexible as we can allow any
efficiency value for a given host halo mass and redshift. We use
this flexibility to explore very different scenarios of galaxy
formation. We present one standard model, which is based on
\citet{DeLucia07} but without an ejected phase or a threshold mass
for star formation. We then introduce four scenarios, each
characterized by a unique feature: zero feedback, cold-accretion in
low mass haloes, stars formed only in merger-induced bursts, and
shutdown of star-formation after mergers. In each model we
tune the other processes so that the resulting galaxies match
observations.
 We do not argue that all these models are
fully plausible, but  show how different processes
like feedback, cooling, and merger-induced bursts can compensate for
each other. This also provides some insight into
the range of efficiencies allowed by observations of
the global properties of the galaxy population over time.
For example, with our Model I, we are able to put a lower limit on the
allowed cooling efficiencies given the general properties of the
galaxy population over time.

We test the five different models presented here against various
observations. We show that all of them reproduce the observed
stellar mass functions at all redshifts reasonably well. We explore
inconsistencies related to the universal SF rate density, and show
the level of agreement that can be achieved between these two
observations. Surprisingly, we find that all our five models reproduce the
observed tilt in the relation of specific SF rates versus stellar mass, unlike
other current SAMs \citep[see e.g.][]{Somerville08,Fontanot09}.
This is likely related to the way that ejection and cooling is
implemented in current SAMs. We plan to investigate this issue in future work.

It is still difficult to assess what it really
means if a given SAM manages to reproduce observations, as
degeneracies are not well understood. How can we know whether or not
a given SAM presents a relatively unique solution to the problem of galaxy
formation and evolution?
We try to improve on these limitations by presenting a
set of models which are very different from each other. This can
help us revisit basic issues of galaxy formation models. For
example, we plan to investigate how galaxy merger-rates behave in
each model, what causes the distinction between red and blue
galaxies, and what shapes the stellar mass function.

We use the set of models developed here to investigate which
observational quantity will break the degeneracy and allow us to
differentiate between the models. The most promising such quantity
is the full distribution of SF rates in different stellar mass bins.
Specifically, SF rates for passive galaxies can give us information
on the physics of SF although they do not affect stellar mass
function, or the cosmic SF density. On the other hand, at low SF rates
it might be that the processes which regulate SF are very different from the
main mode of SF. We find that it is surprisingly difficult for
different models to reproduce the passive fractions of galaxies as a
function of stellar mass, which indicates that this observations may
also be useful for constraining the models. In addition,
cold gas fractions might be useful for discriminating between
different models which all produce a similar galaxy population in
terms of stellar mass and SF. Unfortunately,
up to now published datasets do not take into account
contributions from the atomic and molecular gas components
\citep[for estimates of gas mass ratios see for example][]{Bothwell09,Obreschkow09}.

The formalism presented here can be viewed as an intermediate step
between the halo model and SAMs. This is because all recipes depend
on halo mass only, but the galaxies are followed inside the complex structure
of merger-trees, and their properties will depend
strongly on the halo formation history. Consequently, galaxies which
live inside haloes of the same mass might have very different SF histories.
Similarly to the halo model, our model does not need to be parameterized
a-priori, and the functional dependence of the recipes is obtained
by matching the observations. However, unlike in the halo model, the ingredients
we match (efficiencies of cooling, SF, and feedback) are directly
connected to the underlying physical processes, and we can follow
galaxies in detail through time. This level of modelling is more
complex than the halo model, demands more computational time, and
involves more assumptions.

The current version of our model does not include a computation of
luminosity and color. This limits potential comparisons between our model
galaxies and observations. For example, other SAMs usually compare
the model against luminosity functions, color-magnitude diagrams,
and mass-metallicity relations. On the other hand, comparing to
these observations will require additional model ingredients like dust
obscuration, metallicity enrichment, and the initial stellar mass
function (IMF). It seems that from the above additional
observational measurements, only the mass-metallicity relation is
adding a new constraint to our model. However, the number of
ingredients that would need to be added to the model is quite large, so we suspect that
the level of degeneracy will only increase.
 An additional shortcoming of our model is that it does
not include any morphological information on galaxies. Since the
establishment of the Hubble sequence it is well
known that correlations between morphologies and SF rates do exist
\citep[see a recent work by][]{Schiminovich07}. However, it is unclear if this
correlation is a consequence of the underlying physics, or if it
plays an active role in shaping the SF histories of galaxies as
suggested by \citet{Martig09}. We plan to examine galaxy morphology
in a future work.

We do not attempt to provide improved recipes for SF, cooling or
feedback, even though finding such recipes, which are physically motivated
as well as in agreement with observations both on global and local scales,
is one of the ultimate goals of semi-analytical modelling.
We however believe that our method can help finding such recipes, and studying
the interconnection between them, by opening up the allowed parameter space.

The usual philosophy behind SAMs is that assumptions on the relevant
physical processes are made, and it is then checked whether
the properties of observed galaxies can be reproduced by
tuning a certain set of parameters connected with these physical processes.
We try to solve the inverse problem: how the observed population of
galaxies can constrain the physical mechanisms which drive galaxy formation.
The outcome of our models are average efficiencies per halo mass and redshift,
which need a further interpretation in order to have a physical meaning. For example,
our different results for the cooling efficiencies need to be confronted with
physical models of cooling, in order to find a scenario which is able to predict
such deviations from the standard model. It might be that deviations due to e.g.
metallicity variations and gas density profiles would not be able to give the
required dependencies on halo mass and redshift.
This dependence is more important than our results
on the absolute value of the same quantity, which we have shown
is only poorly constrained by observations.
Lastly, the fact that our recipes do not rely on detailed physical assumptions
means that they generally should not be extrapolated to ranges in halo mass
and redshift for which they have not been tuned. Such problems may
also occur in standard SAMs, although in these cases some guidance for the extrapolation of
each recipe is given.

We note that the cosmological model used here is slightly different
in its parameters from the most recent estimate by
\citet{Komatsu09}. It might be possible to scale the merger-trees
from the Millennium cosmology into the recent cosmology, by changing
the time variable as was suggested by \citet{Neistein09}. If scaling
of the subhalo merger-trees will only require changing the time
variable, then it might be possible to scale our galaxy formation
models by simply transforming only the dependence of the
\emph{recipes} on time. This should leave the results of the models
unchanged. However, while \citet{Neistein09} found that time scaling
should work well for \fof merger-trees, we use
merger-trees based on subhaloes in this work, which might scale
somewhat differently.

We hope that the models developed in this work will help to
interpret both observational data and different galaxy formation
models. To serve this goal
 we provide a public online access to
catalogs of our model galaxies. Our online web
page\footnote{\texttt{http://www.mpa-garching.mpg.de/galform/sesam}} is
also able to run various galaxy formation models with user-defined
parameters over a box size of 62.5 Mpc $h^{-1}$. This volume represents
the results obtained from the full Millennium simulation with a sufficient accuracy
for many applications.
We allow users to upload any kind of recipe which
depends on halo mass and redshift. Results of the model are obtained
within a few seconds and can be downloaded for further analysis.


\section*{Acknowledgments}

We are grateful to Frank van den Bosch for helpful discussions
and for his support in critical points of this project;
to Simon White for reading this paper carefully and for providing
many useful comments;
to Darren Croton for giving an inspiring talk about simple modelling of
galaxy formation;
to Richard Bower, Mike Boylan-Kolchin, Shaun Cole, Niv Drory, Fabio Fontanot,
Marcel Haas, Philip Hopkins, Cheng Li, Umberto Maio,
David Schiminovich, Francesco Shankar, Rob Wiersma and Vivienne Wild for useful discussions;
to Gabriella De Lucia and Qi Guo for allowing us to use their SAM
code, and for many useful discussions;
to Samir Salim for providing us with an electronic version of his data;
to Barbara Catinella for providing us with her results
in advance of publication.
The Millennium Simulation databases used in this paper and the web
application providing online access to them were constructed as part
of the activities of the German Astrophysical Virtual Observatory.
It is a pleasure to thank Gerard Lemson and Laurent Bourges for their work on making our
model publicly available through the internet.
EN is supported by the Minerva fellowship. SW is supported by the
German-Israeli Foundation (GIF).

\bibliographystyle{mn2e}
\bibliography{ref_list}


\section*{Appendix A: Efficiencies values}

\label{sec:app_effic}

In tables \ref{tab:coll_effic_I} -- \ref{tab:sf_effic_I} we provide numerical values for the
cooling and SF efficiencies used by our models. We use bilinear interpolation to sample the
values into a fine grid which is then used by our code.

\begin{table}
\caption{Values of cooling efficiencies, $\log \fc$, where $\fc$ is in units
of Gyr$^{-1}$. The values shown here are for Model I, and are plotted in Fig.~\ref{fig:cool_effic}.
Halo mass is in units of $\hmsun$ and is shown at the left column, time is in Gyr.}
\begin{center}
\begin{tabular}{l || c | c | c | c | c | c }
   Halo  & $t=0.80$  & 2.24 & 3.38 & 5.97 & 10.27 & 13.58     \\
 mass     &  $\z=7$  &  3.0 &   2.0  &  1.0 &   0.3  &   0  \\
\hline
\hline
  10.25 &  -2.10 &  -2.50 &  -2.80 &  -2.80 &  -2.80 &  -2.80 \\
 10.75 &  -1.90 &  -2.40 &  -2.40 &  -2.40 &  -2.40 &  -2.40 \\
 11.00 &  -0.90 &  -1.30 &  -1.90 &  -2.10 &  -2.10 &  -2.10 \\
 11.25 &  -0.20 &  -0.90 &  -1.50 &  -1.60 &  -1.80 &  -2.00 \\
 11.50 &  0.10 &  -0.70 &  -1.00 &  -1.40 &  -1.70 &  -1.80 \\
 11.75 &  0.30 &  -0.50 &  -0.70 &  -1.30 &  -1.70 &  -1.80 \\
 12.00 &  0.10 &  -0.70 &  -0.70 &  -1.30 &  -1.80 &  -2.00 \\
 12.25 &  -0.20 &  -0.70 &  -1.30 &  -1.50 &  -2.00 &  -4.00 \\
 12.50 &  -0.70 &  -1.70 &  -2.00 &  -4.00 &  -4.00 &  -4.00 \\
\end{tabular}
\end{center}
\label{tab:coll_effic_I}
\end{table}

\begin{table}
\caption{Same as table \ref{tab:coll_effic_I}, but for Model II.}
\begin{center}
\begin{tabular}{l || c | c | c | c | c | c }
  Halo  & $t=0.80$  & 2.24 & 3.38 & 5.97 & 10.27 & 13.58     \\
 mass     &  $\z=7$  &  3.0 &   2.0  &  1.0 &   0.3  &   0  \\
\hline
\hline
10.25 &  -2.00 &  -2.00 &  -2.20 &  -2.20 &  -2.40 &  -2.60 \\
 10.75 &  -1.20 &  -1.20 &  -1.40 &  -1.60 &  -2.20 &  -2.40 \\
 11.00 &  -0.40 &  -0.60 &  -1.10 &  -1.30 &  -1.80 &  -2.10 \\
 11.25 &  -0.10 &  -0.40 &  -0.70 &  -0.80 &  -1.30 &  -1.80 \\
 11.50 &  0.50 &  0.20 &  -0.20 &  -0.70 &  -1.00 &  -1.50 \\
 11.75 &  0.70 &  0.40 &  -0.10 &  -0.50 &  -1.00 &  -1.40 \\
 12.00 &  0.80 &  0.47 &  0.10 &  -0.50 &  -1.00 &  -1.50 \\
 12.15 &  0.80 &  0.47 &  0.10 &  -0.50 &  -1.00 &  -2.50 \\
 12.30 &  0.80 &  0.40 &  -0.20 &  -2.00 &  -4.00 &  -4.00 \\
 12.50 &  0.80 &  0.20 &  -2.00 &  -4.00 &  -4.00 &  -4.00 \\
\end{tabular}
\end{center}
\label{tab:coll_effic_II}
\end{table}

\begin{table}
\caption{Same as table \ref{tab:coll_effic_I}, but for Models III \& IV.}
\begin{center}
\begin{tabular}{l || c | c | c | c | c | c }
  Halo   & $t=0.80$  & 2.24 & 3.38 & 5.97 & 10.27 & 13.58     \\
 mass     &  $\z=7$  &  3.0 &   2.0  &  1.0 &   0.3  &   0  \\
\hline
\hline
 10.25 &  0.92 &  0.47 &  0.29 &  0.05 &  -0.19 &  -0.31 \\
 10.75 &  0.92 &  0.47 &  0.29 &  0.05 &  -0.19 &  -0.31 \\
 11.00 &  0.92 &  0.47 &  0.29 &  0.05 &  -0.19 &  -0.31 \\
 11.25 &  0.92 &  0.47 &  0.29 &  0.05 &  -0.19 &  -0.80 \\
 11.50 &  0.92 &  0.47 &  0.29 &  -0.50 &  -0.70 &  -2.00 \\
 11.75 &  0.92 &  -0.10 &  -0.20 &  -1.00 &  -1.50 &  -4.00 \\
 12.00 &  0.50 &  -0.50 &  -0.70 &  -4.00 &  -4.00 &  -4.00 \\
 12.15 &  -1.00 &  -1.00 &  -1.00 &  -4.00 &  -4.00 &  -4.00 \\
 12.30 &  -4.00 &  -4.00 &  -4.00 &  -4.00 &  -4.00 &  -4.00 \\
 12.50 &  -4.00 &  -4.00 &  -4.00 &  -4.00 &  -4.00 &  -4.00 \\
\end{tabular}
\end{center}
\label{tab:coll_effic_III}
\end{table}

\begin{table}
\caption{Same as table \ref{tab:coll_effic_I}, but for Model V.}
\begin{center}
\begin{tabular}{l || c | c | c | c | c | c }
  Halo   & $t=0.80$  & 2.24 & 3.38 & 5.97 & 10.27 & 13.58     \\
 mass     &  $\z=7$  &  3.0 &   2.0  &  1.0 &   0.3  &   0  \\
\hline
\hline
 10.25 &  -2.00 &  -2.00 &  -2.20 &  -2.20 &  -2.40 &  -2.60 \\
 10.75 &  -1.20 &  -1.20 &  -1.40 &  -1.60 &  -2.20 &  -2.40 \\
 11.00 &  -0.40 &  -0.60 &  -1.10 &  -1.30 &  -1.80 &  -2.10 \\
 11.25 &  -0.10 &  -0.40 &  -0.70 &  -0.80 &  -1.30 &  -1.80 \\
 11.50 &  0.50 &  0.20 &  -0.20 &  -0.70 &  -1.00 &  -1.50 \\
 11.75 &  0.70 &  0.40 &  -0.10 &  -0.50 &  -1.00 &  -1.40 \\
 12.00 &  0.80 &  0.47 &  0.10 &  -0.50 &  -1.00 &  -1.50 \\
 12.15 &  0.80 &  0.47 &  0.10 &  -0.50 &  -1.00 &  -1.50 \\
 12.30 &  0.80 &  0.40 &  -0.20 &  -1.50 &  -1.50 &  -1.50 \\
 12.50 &  0.80 &  0.20 &  -1.50 &  -1.50 &  -1.50 &  -1.50 \\
\end{tabular}
\end{center}
\label{tab:coll_effic_V}
\end{table}

\begin{table}
\caption{Same as table \ref{tab:coll_effic_I}, but for the model DLB07.
Here we sample only few points out of the data presented in Fig.~\ref{fig:cool_effic}. }
\begin{center}
\begin{tabular}{l || c | c | c | c | c | c }
  Halo   & $t=0.80$  & 2.24 & 3.38 & 5.97 & 10.27 & 13.1     \\
 mass     &  $\z=7$  &  3.0 &   2.0  &  1.0 &   0.3  &   0  \\
\hline
\hline
 10.25 &  1.51 &  0.98 &  0.67 &  -0.10 &  -0.36 &  -0.43 \\
 10.75 &  1.29 &  0.83 &  0.69 &  -0.06 &  -0.38 &  -0.53 \\
 11.00 &  1.15 &  0.35 &  0.19 &  0.53 &  -0.32 &  -0.49 \\
 11.25 &  0.76 &  0.33 &  0.17 &  -0.06 &  -0.28 &  -0.44 \\
 11.50 &  0.69 &  0.25 &  0.09 &  -0.10 &  -0.31 &  -0.37 \\
 11.75 &  0.58 &  0.21 &  -0.00 &  -0.26 &  -0.43 &  -0.50 \\
 12.00 &  -2.29 &  0.03 &  -0.10 &  -0.45 &  -0.78 &  -0.96 \\
 12.25 &  -1.05 &  -0.28 &  -0.54 &  -0.89 &  -4.00 &  -4.00 \\
 12.50 &  -2.39 &  -0.58 &  -1.33 &  -4.00 &  -4.00 &  -4.00 \\
\end{tabular}
\end{center}
\label{tab:coll_effic_DLB07}
\end{table}

\begin{table}
\caption{Values of SF efficiencies, $\log \fs$, where $\fs$ is in units
of Gyr$^{-1}$. The values shown here are for Model I, and are plotted in Fig.~\ref{fig:sf_effic}.
Halo mass is in units of $\hmsun$ and is shown at the left column, time is in Gyr.}
\begin{center}
\begin{tabular}{l || c | c | c | c | c | c }
  Halo   & $t=0.80$  & 2.24 & 3.38 & 5.97 & 10.27 & 13.58     \\
 mass     &  $\z=7$  &  3.0 &   2.0  &  1.0 &   0.3  &   0  \\
\hline
\hline
 10.25 &  1.00 &  0.70 &  0.50 &  0.10 &  -0.90 &  -0.90 \\
 10.75 &  1.00 &  0.70 &  0.50 &  0.10 &  -0.90 &  -0.90 \\
 11.00 &  1.00 &  0.70 &  0.60 &  0.20 &  -0.50 &  -0.70 \\
 11.25 &  1.00 &  0.70 &  0.70 &  0.30 &  -0.20 &  -0.70 \\
 11.50 &  1.00 &  0.75 &  0.80 &  0.40 &  -0.20 &  -0.80 \\
 11.75 &  1.00 &  0.75 &  0.70 &  0.20 &  -0.50 &  -0.90 \\
 12.00 &  1.00 &  0.75 &  0.70 &  -0.70 &  -1.80 &  -2.50 \\
 12.25 &  1.00 &  0.75 &  -0.10 &  -1.00 &  -2.00 &  -2.70 \\
 12.50 &  1.00 &  0.75 &  -0.70 &  -2.50 &  -3.00 &  -3.00 \\
\end{tabular}
\end{center}
\label{tab:sf_effic_I}
\end{table}
%


\section*{Appendix B: Additional data -- Specific SF rates}
\label{sec:app_ssfr}

In this section we provide a comparison of our models SF rates at $\z=0$ to the
observational study by \citet{Salim07}.

\begin{figure*}
\centerline{\psfig{file=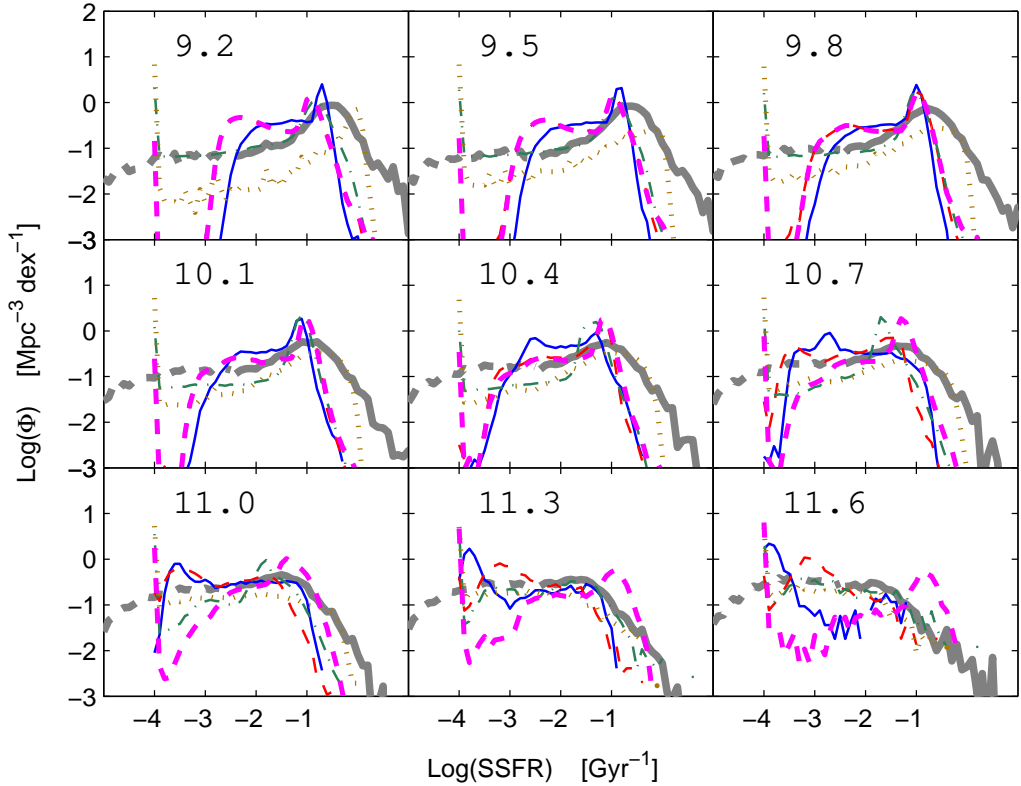,width=150mm,bbllx=30mm,bblly=80mm,bburx=188mm,bbury=200mm,clip=}}
\caption{The probability  distribution of specific SF rates for
different bins in stellar mass at $\z=0$. The figure is similar to
Fig.~\ref{fig:ssfr_cuts} but it compares to the observational data
of \citet{Salim07}.} \label{fig:ssfr_cuts_salim}
\end{figure*}


\label{lastpage}

\end{document}